\documentclass[aps,prb,twocolumn,floatfix]{revtex4-2}
\usepackage[colorlinks=true,citecolor=blue,linkcolor=blue,urlcolor=blue]{hyperref}

\usepackage[sort&compress]{natbib}
\usepackage{graphicx,times}
\usepackage{color} 
\usepackage{sansmath}
\usepackage{amssymb}

\usepackage{subeqnarray}
\usepackage{diagbox}
\usepackage{makecell}
\usepackage{amsmath}
\usepackage{mathtools}
\usepackage{empheq}
\usepackage{bbding}

\newcommand{\msf}[1]{\mathsf{#1}}
\newcommand{\mcl}[1]{\mathcal{#1}}
\newcommand{\bsb}[1]{\boldsymbol{#1}}

\newcommand{\mathbbm}[1]{\text{\usefont{U}{bbm}{m}{n}#1}}

\DeclareSymbolFont{greekletters}{OML}{FiraSans}{m}{n}

\DeclareMathOperator{\msOmega}{\msf{\Omega}}
\DeclareMathOperator{\msJ}{\msf{J}}
\DeclareMathOperator{\msH}{\msf{H}}
\DeclareMathOperator{\msJH}{\msf{JH}}

\DeclareMathOperator{\mLambda}{\msf{\Lambda}}

\DeclareMathOperator{\bv}{\bsb{v}}

\DeclareMathOperator{\bphi}{\bsb{\phi}}
\DeclareMathOperator{\bPhi}{\bsb{\Phi}}
\DeclareMathOperator{\bQ}{\bsb{Q}}
\DeclareMathOperator{\bPi}{\bsb{\Pi}}
\DeclareMathOperator{\bpi}{\bsb{\pi}}
\DeclareMathOperator{\bP}{\bsb{P}}
\DeclareMathOperator{\bp}{\bsb{p}}

\DeclareMathOperator{\bx}{\bsb{x}}
\DeclareMathOperator{\by}{\bsb{y}}
\DeclareMathOperator{\bz}{\bsb{z}}
\DeclareMathOperator{\bX}{\bsb{X}}
\DeclareMathOperator{\bY}{\bsb{Y}}
\DeclareMathOperator{\bZ}{\bsb{Z}}
\DeclareMathOperator{\be}{\bsb{e}}

\DeclareMathOperator{\bq}{\bsb{q}}

\DeclareMathOperator{\bPsi}{\mathbf{\Psi}}

\DeclareMathOperator{\msB}{\msf{B}}
\DeclareMathOperator{\msC}{\msf{C}}
\DeclareMathOperator{\msD}{\msf{D}}

\DeclareMathOperator{\msY}{\msf{Y}}
\DeclareMathOperator{\msW}{\msf{W}}
\DeclareMathOperator{\msZ}{\msf{Z}}
\DeclareMathOperator{\msS}{\msf{S}}
\DeclareMathOperator{\msF}{\msf{F}}
\DeclareMathOperator{\msL}{\msf{L}}
\DeclareMathOperator{\msG}{\msf{G}}

\DeclareMathOperator{\msP}{\msf{P}}

\DeclareMathOperator{\msM}{\msf{M}}
\DeclareMathOperator{\msN}{\msf{N}}

\newcommand{\mone}{\mathbbm{1}}

\usepackage{svg}

\newcommand{\orcid}[1]{\href{https://orcid.org/#1}{\includegraphics[width=8pt]{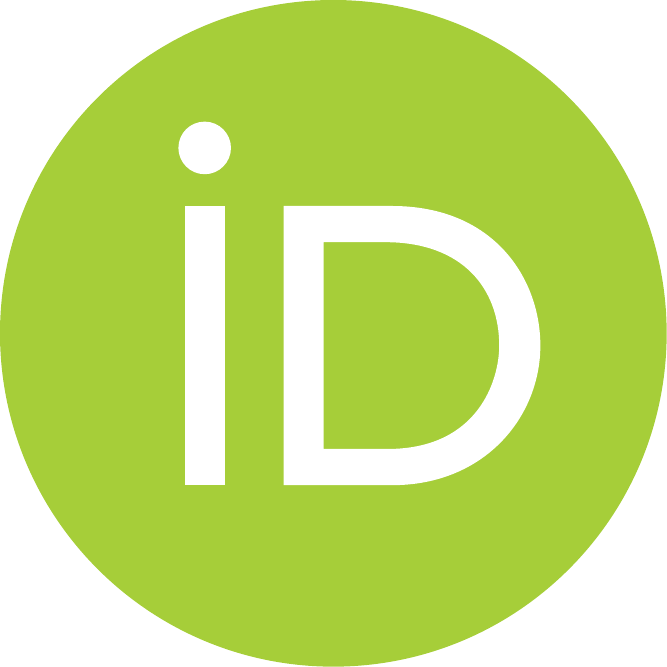}}}

\begin{document}
\title{Algebraic canonical quantization of lumped superconducting networks}
\author{I. L. Egusquiza\,\orcid{0000-0002-5827-8027}}
\email{inigo.egusquiza@ehu.es \Envelope}
\affiliation{Department of Physics, University of the Basque Country UPV/EHU, Apartado 644, 48080 Bilbao, Spain}
\affiliation{EHU Quantum Centre, University of the Basque Country UPV/EHU, Apartado 644, 48080 Bilbao, Spain}
\author{A. Parra-Rodriguez\,\orcid{0000-0002-0896-9452}}
\email{ adrian.parra.rodriguez@gmail.com \Envelope}
\affiliation{Department of Physics, University of the Basque Country UPV/EHU, Apartado 644, 48080 Bilbao, Spain}
\affiliation{Institut Quantique and Département de Physique, Université de Sherbrooke, Sherbrooke, Qu\'ebec J1K 2R1, Canada}

\begin{abstract}
We present a systematic canonical quantization procedure for  lumped-element superconducting networks by using a redundant configuration-space description. The algorithm is based on an original, explicit, and constructive implementation of the symplectic diagonalization of positive semidefinite Hamiltonian matrices,  a particular instance of Williamson's theorem. With it, we derive canonically quantized discrete-variable descriptions of passive causal systems. We exemplify the algorithm with representative {\it singular} electrical networks, a nonreciprocal extension for the black-box quantization method, as well as an archetypal Landau quantization problem.
\end{abstract}

\maketitle
\section{Introduction}
Superconducting circuits \cite{Devoret:2013} are the current leading technology for quantum information processing \cite{Arute:2019,Martinis:2020,Wu:2021}. Complex as their quantum dynamics may be, simple effective models based on microwave circuit theory have  been extremely succesful in explaining collective quantum phenemena therein. Indeed, the 3-dimensional (3D) Maxwell-London theory together with the Josephson equations~\cite{Josephson:1962} are the main phenomenological framework routinely used to describe them~\cite{Tinkham:2004,Jackson:1999}. However, in certain cases where characteristic lengths are smaller than the wavelengths involved, e.g., when energy is localised at very small volumes, such 3D equations can be further simplified to Kirchhoff's laws, which are representable through ideal lumped-elements circuits. In fact, the canonical quantization is that of the differential equations governing the dynamics of the collective degrees of freedom of those elements (like capacitors, inductors and Josephson junctions)~\cite{YurkeDenker:1984,Devoret:1997,Vool:2017}. 

Nevertheless, bigger electromagnetic environments, such as superconducting 3D cavities, are still commonly used in experiments to achieve, for instance, long coherence times~\cite{Paik:2011}. In order to incorporate these, and more general linear passive systems with well-defined probing ports in  Kirchhoff's framework, a two tier process has been suggested. First, a classical multiphysics analysis of their linear response is performed, and then, its output is fed to classical engineering theorems for fraction expansions of causal passive matrices~\cite{Newcomb:1966}. In this way, it is possible to obtain lumped-element networks of the low-lying energy spectrum to the required level of accuracy which are fit for quantization, a routine better known as the {\it black-box} approach~\cite{Nigg:2012,Solgun:2014,Solgun:2015}. The classical expansions can definitely include effective breaking of time-reversal symmetry~\cite{Koch:2010} through ideal gyrator (2-port) or circulator (N-port) devices~\cite{Sliwa:2015,Chapman:2017,Mueller:2018,Kerckhoff:2015,Viola:2014,Mahoney:2017,Barzanjeh:2017}. These are nonreciprocal (NR) elements that can be used for noise isolation or signal routing~\cite{Tellegen:1948}, and that implement flux-charge constraints. Necessarily, they have to be studied in the quantum regime as well~\cite{ParraRodriguez:2019,Rymarz:2018}. Furthermore, in order to represent general linear responses~\cite{Newcomb:1966} the introduction of  Belevitch transformers is required~\cite{Belevitch:1950}, yet another type of multi-terminal element, which provide direct constraints between fluxes, and between charges. 

Thus, it has become of great interest to find a systematic quantum circuit analysis~\cite{YurkeDenker:1984,Chakravarty:1986,Yurke:1987,Werner:1991,Devoret:1997,Paladino:2003,Burkard:2004,Burkard:2005,Bourassa:2012,Nigg:2012,Solgun:2014,Solgun:2015,Ulrich:2016,Mortensen:2016,Malekakhlagh:2016,Vool:2017,Malekakhlagh:2017,Gely:2017,ParraRodriguez:2018,ParraRodriguez:2019,Gely:2020,Minev:2020,Minev:2021,ParraRodriguez:2022,Mariantoni:2020,Minev:2021,Chitta:2022,Rajabzadeh:2022} of electrical networks that treats all linear passive lumped elements (capacitors, inductors, ideal nonreciprocal elements and Belevitch transformers) on an equal footing. This analysis must deal with obstacles such as singular kinetic matrices that would prevent straightforward Legendre transformations. We stress a Lagrangian formulation since it gives us both canonical Poisson brackets, later to be promoted to commutators, and Hamiltonian functions on phase space, later to be promoted to operators.

In this article, we present a systematic procedure for deriving Hamiltonians together with canonical quantizations of lossless lumped-element networks. This procedure is based on a Lagrangian written in terms of an enlarged configuration space with both flux and charges variables \cite{Jeltsema:2009,Ulrich:2016,ParraRodriguez:2022}. Such enlargement with respect to most of  previous approaches (where Lagrangians were written in either flux or charge variables) allows us to replace the troublesome Legendre transformation for singular kinetic matrices by a simple nonsingular matrix inversion. Naturally, there is a cost to pay with this approach, as redundancies will be certainly introduced in the form of {\it free particles} and {\it nondynamical} degrees of freedom. However, making use of the full-fledged Williamson theorem for normal forms of quadratic systems~\cite{Williamson:1936}, these subsectors can be systematically disentagled from the nontrivial dynamical sector and canonical quantization can follow effortlessly. 

The underlying mathematical  keystone  of the present work is a simple and constructive algorithm to find systematically  the symplectic normal forms of positive semi-definite symmetric (real) matrices (with eigenvalues $\lambda_i\geq 0$) inspired by H\"ormander's analysis~\cite{Hoermander:1995} of Williamson's seminal article \cite{Williamson:1936}. We consider and propose the application of this  routine to find canonical quantum descriptions of general passive causal systems presenting a discrete set of frequencies. We stress that this includes the  possibility of constraints that hitherto have been considered an obstacle to achieve that goal.  In particular, we exploit it to obtain systematically exact quantum descriptions of superconducting lumped-element networks, including cases analogous to the Landau problem, with discrete macroscopic conjugated charges and fluxes replacing microscopic position variables. 

The results  obtained here generalize systematic derivations of canonically quantized Hamiltonians for reciprocal ~\cite{Devoret:1997,Vool:2017,Burkard:2004,Burkard:2005,Ulrich:2016,Hassler:2019}, and nonreciprocal~\cite{ParraRodriguez:2019,Rymarz:2018} superconducting lumped-element networks based on graph theory, and will provide support for more general quasi-lumped descriptions of circuits~\cite{ParraRodriguez:2018,Minev:2021,ParraRodriguezPhD:2021}. We remark that a particular and important consequence of the application of our algorithm is the generalization of the black-box quantization methods of non-linear degrees of freedom coupled to nonreciprocal linear systems~\cite{Nigg:2012,Solgun:2014,Solgun:2015}.

The article is structured as follows: in Sec.~\ref{Sec:Williamson} we introduce an explicit constructive algorithm of Williamson's theorem to find symplectic transformations that diagonalize real positive-semidefinite matrices of dimension $2n$. For illustration in a simple understandable case, we apply it to a Landau quantization problem with two linearly-coupled charged particles in a homogeneous magnetic field. We make full use of the main theorem in Sec.~\ref{Sec:QuantizationLumpedNetworks} to put forward a systematic process to find canonically quantized Hamiltonians of lumped-element superconducting networks described in terms of Kirchhoff's equations, starting from a Lagrangian written in the enlarged configuration space.  In Sec.~\ref{Sec:JJs_Y_NR_BB}, we implement the algorithm to quantize a {\it singular} nonreciprocal circuit containing both nondynamical degrees of freedom and free particles using both (i) a black-box approach and (ii) keeping the full cosine potential of the nonlinear degrees of freedom. We finally draw some conclusions and describe possible additional applications in Sec.~\ref{Sec:Conclusions}.

\section{Williamson's Theorem for Positive Semi-definite Symmetric Matrices}
\label{Sec:Williamson}
The study  of the classification, with canonical presentations, of quadratic forms has a long history. For instance, Weierstrass (\cite{Weierstrass:1858} in p. 233 of \cite{Weierstrass:1894},  and \cite{Weierstrass:1868} published with some changes in p. 19 of \cite{Weierstrass:1895}) looked into the conditions for pairs of quadratic forms to be put in canonical form (Gestalt). Part of the motivation of Weierstrass was the application of these canonical forms to the theory of small oscillations, and this reason and others kept the topic alive. In 1936, Williamson gave the complete classification of quadratic forms under symplectic transformations \cite{Williamson:1936}, over general fields of characteristic zero. His classical work did not provide us with explicit constructive methods nor did it include a complete catalogue of canonical forms.  Laub and Meyer \cite{Laub:1974} did provide a listing of canonical forms, which was refined and completed by H\"ormander \cite{Hoermander:1995}. In fact, many other authors have looked into the issue without being aware of each other \cite{Bryuno_1988,Churchill:1999}. An alternative source of information for physicists is the compilation of symplectic normal forms for quadratic Hamiltonians presented by Arnol'd in Appendix 6 of \cite{Arnold:2013} and attributed by him to D. M. Galin, without further reference. 
	
	In a more recent development, the study of quantum information with continuous variables, and in particular of Gaussian states \cite{Adesso:2014}  has brought a special case of Williamson's theorem to the fore, that of positive definite quadratic matrices. Of special relevance in this context is the symplectic diagonalization of covariance matrices \cite{Pirandola:2009}. Some attention to the topic has also been given in the context of superconducting circuits and other quantum systems. In these applications a constructive approach is required, beyond the descriptions in the theorems, and much has been made of this fact for the simplest case of positive definite Hamiltonians ~\cite{Burgoyne:1974}, while that of positive semidefinite Hamiltonians has been referred to but not addressed. Here we remedy that deficiency.
	
Let us denote as $\mathsf{H}$ the matrix corresponding to a quadratic Hamiltonian $H=\bx^T\mathsf{H}\bx/2$, while $\mathsf{J}$ is the canonical symplectic matrix, such that the (classical and Heisenberg) dynamics reads
\begin{align}
	\dot{\bx}=\msJ\msH\bx\label{eq:eoms_phase_space_H}
\end{align}
for phase space coordinates $\bx=(q_1,..., q_n, p_1,..., p_n)^T=(x_1,...,x_{2n})^T$.  The content of Williamson's theorem for positive semidefinite Hamiltonians is that there are only three possiblities for independent degrees of freedom  present in Eq. \eqref{eq:eoms_phase_space_H} when $\msH$ belongs to that class~\cite{Hoermander:1995}:  (i) harmonic oscillators, evolving as  $\ddot{\by}_{\mathrm{ho}}=\msOmega^2\by_{\mathrm{ho}}$, where $\msOmega$ is a positive definite diagonal matrix of frequencies, (ii) free particles with  $\ddot{\by}_{\mathrm{fp}}=0$, and (iii) nondynamical variables, i.e.,  $\by_{\mathrm{nd}}(t)=\by_{\mathrm{nd}}(0)$. These variables $\by$ are the desired objective, related to the initial ones by a symplectic transformation $\bx=\msS\by$.
	
	To proceed, we have to identify the linear subspaces $K_1=\mathrm{ker}\left[\mathsf{JH}\right]= \mathrm{ker}\left[\mathsf{H}\right]$ and $K_2=\mathrm{ker}\left[\left(\mathsf{JH}\right)^2\right]$. If $K_1=K_2$,  there are no free particles, and all the  variables corresponding to $K_1$ are nondynamical. More generally, $n_{\mathrm{f}}=\mathrm{dim}(K_2)-\mathrm{dim}(K_1)$ is the number of free particles and $\mathrm{dim}(K_1)-n_f=2n_{\mathrm{nd}}$ is even and twice the number of nondynamical degrees of freedom $n_{\mathrm{nd}}$. Notice that $\mathrm{dim}(K_2)=2 n_{\mathrm{f}}+2n_{\mathrm{nd}}$ is necessarily even. There are $n_{\mathrm{ho}}=n-n_{\mathrm{f}}-n_{\mathrm{nd}}$ harmonic oscillators, to complete the description, with $2n$ the dimension of phase space.

	The classification  above translates into the existence of  a change of basis matrix $\mathsf{F}$ for which
	\begin{align}
		\label{eq:jordanform}
		\mathsf{F}^{-1} \mathsf{JH} \mathsf{F} = \msD_{\msJ} =
		\begin{pmatrix}
			0&0&0&0&0&0\\ 0&0&0&0&0&0\\ 0&0& 0&\mone&0&0\\ 0&0&0&0&0&0 \\ 0&0&0&0& i \msOmega&0\\  0&0&0&0&0&- i \msOmega
		\end{pmatrix}.
	\end{align}
	The order has been chosen 
	in the sequence nondynamical/free particles/harmonic oscillators for the degrees of freedom. The dimensions, not explicit in the expression above, run in the sequence $\left\{n_{\mathrm{nd}},n_{\mathrm{nd}},n_{\mathrm{f}},n_{\mathrm{f}},n_{\mathrm{ho}},n_{\mathrm{ho}},\right\}$ for the diagonal boxes. This matrix $\mathsf{F}$ is the output of the usual Jordan canonical form process for $\mathsf{JH}$, up to ordering.
	
	Generically, this change of basis matrix does not diagonalize $\mathsf{H}$ by a symplectic transformation. For starters, it is necessarily complex. However, an adequate transformation will be of the form 
	\begin{equation}
		\label{eq:Sfi}
		\mathsf{S}=\msF\mLambda\msF_0^{-1}\,,
	\end{equation}
	where $\mathsf{F}_0$ is the matrix
	\begin{equation}
		\label{eq:f0}
		\mathsf{F}_0=
		\begin{pmatrix}
			\mone&0&0&0&0&0\\0&\mone&0&0&0&0\\0&0&\mone&0&0&0\\0&0&0&\mone&0&0\\0&0&0&0&\mone&\mone\\0&0&0&0&i\mone&-i\mone
		\end{pmatrix}\,.
              \end{equation}
              As we show explicitly in Appendix \ref{sec:sympl-diag-posit}, the structure of $\mathsf{F}_0$ comes about because the real and imaginary parts of the eigenvectors of $\mathsf{JH}$ with purely imaginary eigenvalues can be arranged in symplectic pairs, while the generalized eigenvectors of the zero eigenvalue are real.
              
              To complete the description, $\mathsf{\Lambda}$ is a block diagonal matrix, whose detailed structure depends on the existence or not of degeneracies both in the harmonic oscillator sector and in the generalized eigenspace of $\mathsf{JH}$ for eigenvalue 0. Let us first examine its structure in the harmonic oscillator sector.  If there are no nonzero degenerate frequencies, the corresponding block of $\mathsf{\Lambda}$ consists of two diagonal matrices of dimension $n_{\mathrm{ho}}$ for the harmonic oscillator sector, with the diagonal elements being normalization factors. This is in fact the case that has been frequently explored in the literature, since it pertains to the positive definite case without degeneracies, the only one that has merited full attention, as it is the easiest. Those diagonal elements  in this case simply give  the normalization of each eigenvector of $\mathsf{JH}$ for purely imaginary eigenvalue, such that its real and imaginary part are symplectic orthonormal.

              Were there a degeneracy in the nonzero spectrum of $\mathsf{JH}$, the corresponding block for $\mathsf{\Lambda}$ will be due to symplectic Gram--Schmidt orthonormalization of the real and imaginary parts of the eigenvectors for that eigenvalue. Similarly, the $\mathsf{\Lambda}$ is required, in the free particle and non-dynamical sector, to ensure that the free particle sector is symplectic orthogonal to the non-dynamical one, and that the resultant free particle sector is symplectic orthonormal. This can be done explicitly, as shown in Appendix \ref{sec:sympl-diag-posit}, or through the condition that 
\begin{equation}
		\label{eq:condonS}
		\mathsf{J} =\mathsf{S}^T
		\begin{pmatrix}
			\omega_0&0\\0& \mathsf{J}_{\mathrm{ho}}
		\end{pmatrix}\mathsf{S}\,,
	\end{equation}
	where $\omega_0$ is an antisymmetric $(2n_{\mathrm{nd}}+2n_{\mathrm{f}})\times(2n_{\mathrm{nd}}+2n_{\mathrm{f}})$ while $\mathsf{J}_{\mathrm{ho}}$ is the canonical symplectic matrix, now for the harmonic oscillator sector. If there are degeneracies, the harmonic oscillator sector of $\mathsf{\Lambda}$ becomes block diagonal, with pairs of invertible matrices for each degeneracy subspace. Complete details are presented in Appendix \ref{sec:sympl-diag-posit}.  Once $\mathsf{\Lambda}$ has been determined, we compute $\mathsf{S}^T\mathsf{H}\mathsf{S}$, resulting in a block diagonal matrix that is in fact canonically diagonal in the harmonic oscillator sector, i.e. $\mathrm{diag}\left(\mathsf{\Omega},\mathsf{\Omega}\right)$. If one desires full symplectic diagonalization an additional symplectic Gram--Schmidt process will be required for $K_2$, as made explicit in  Appendix \ref{sec:sympl-diag-posit}.
	
\subsection{A familiar example: the Landau problem}
\label{Sec:Landau levels}
\begin{figure}[h]
	\centering
	\includegraphics[width=0.25\textwidth]{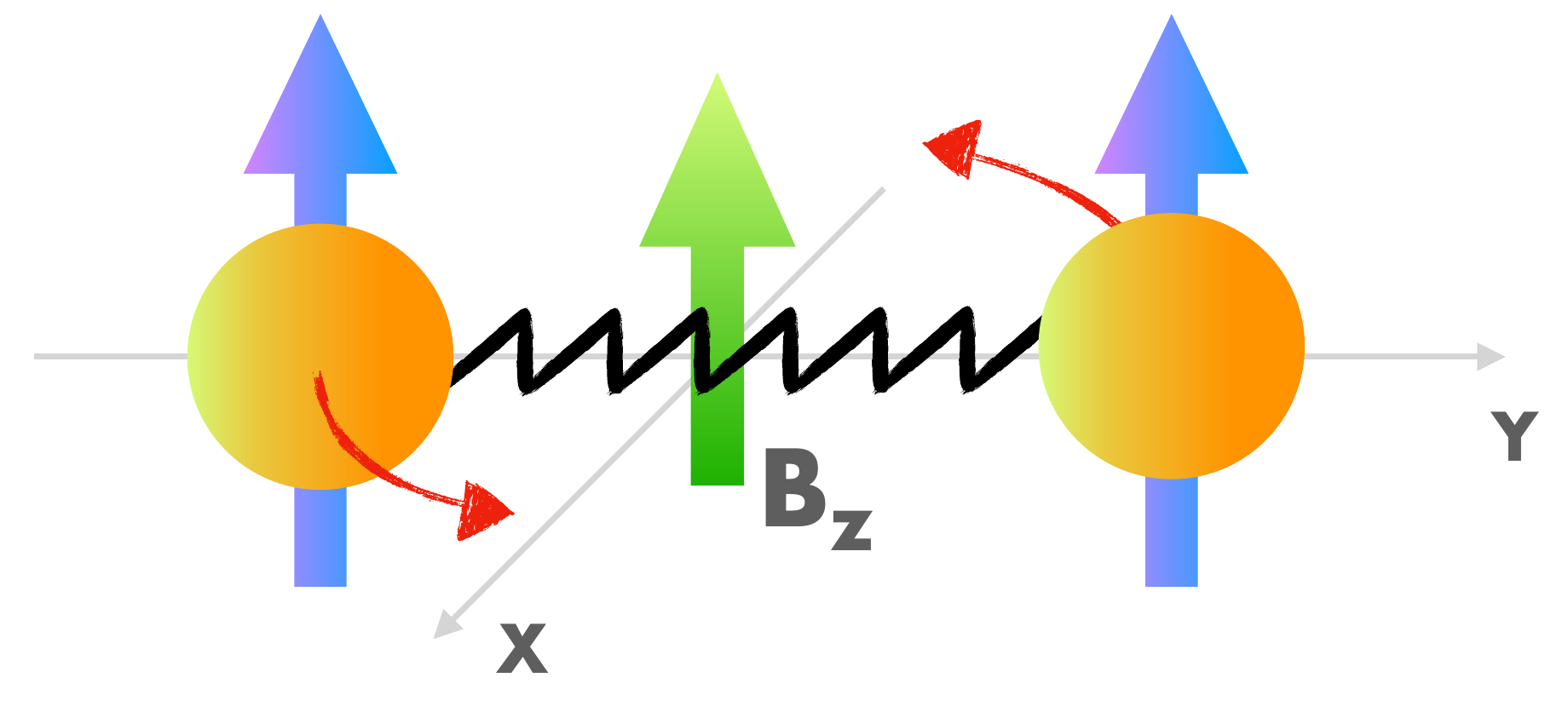}
	\caption{Two coupled charged particles subjected to the action of a magnetic field. The dynamics of this system naturally separate in two orthogonal sectors, the in-plane ($x$--$y$) and normal ($z$) ones.}
	\label{fig:Landau_fig}
\end{figure}

The Landau problem of one charged particle moving in a homogeneous magnetic field is very well known indeed. Simple as its statement is, its correct solution requires a canonical transformation that involves both positions and momenta. Furthermore, it is the starting point for our understanding of the quantum Hall effect. We consider a slight extension of the Landau problem, and address it with the symplectic diagonalization  algorithm  so as to illustrate its applicability in a simple example. Let us thus consider a pair of identical linearly-coupled charged particles subjected to a constant and homogeneous external magnetic field in the $z$-direction ($\bsb{B}=(0,0,B)$), see Fig.\ref{fig:Landau_fig}. 

Working in the symmetric gauge, i.e., $\bsb{A}(\bq_i)=-\frac{1}{2}\left(\bsb{B}\times \bq_i\right)$, a redundant  minimal-coupling Hamiltonian can be written as 
\begin{align}\label{eq:MinimalCoupling}
	H=&\sum_i\frac{\left(\bp_i-e\bsb{A}(\bq_i)\right)^2}{2m}+\frac{k}{2}(\bq_1-\bq_2)^2,
\end{align}
where the dynamics of the phase-space separates in two independent subsectors, $2H=\bX_z^T\msH_z\bX_z+\bX_{xy}^T\msH_{xy}\bX_{xy}$, i.e. vertical and in-plane motion. Williamson's analysis of $\msH_z$ yields, as symplectic diagonal degrees of freedom, one harmonic oscillator and one free particle. In the limit of the coupling constant going to zero ($k\rightarrow0$), obviously, the harmonic oscillator frequency tends to zero and in fact we have two free particles.

The analogous analysis in the $x$--$y$ plane shows the existance of three harmonic conjugate pairs, and an extra non-dynamical one. Again, by taking the zero-coupling limit, the dynamics in the plane split in two pairs of harmonic and nondynamical conjugated variables. See Appendix~\ref{App:Landau_quantization} for further details and an explicit expression for a symplectic diagonalizing matrix.

Observe that the procedure by itself does not prescribe a meaning for the nondynamical sector. Thus, in the case of the mechanical system with just these two distinguishable particles, the nondynamical sector is set to being identically zero, and would not be detectable if the system were realized. Not so if we take this study as part of  second quantization for fermions, say, in which case we would understand  fermions in the same  dynamical state to be allowed to be in different states in the nondynamical sector, as is done in the analysis of the quantum Hall effect. Nonetheless, our procedure allows us the systematic identification of the different sectors.

In this particular case, one might wonder whether our analysis depends in any way on having chosen the symmetric gauge. However, one crucial point to be stressed in the Williamson diagonalization procedure is that the canonical form under symplectic transformations is a symplectic invariant, by construction. And since all the gauge transformations that preserve the quadratic character of the Hamiltonian are actually implementable as symplectic transformations, the physical content of the model is seen as clearly gauge invariant, and achieved systematically. The gauge invariance of the problem has led to some confusion in the literature as to the conserved quantities associated with the model. Again, looking at the issue from the point of view of symplectic invariance allows the immediate  identification of the maximum number of linear and quadratic conserved quantities for these families of systems. Their physical interpretation is on the other hand directly dependent on how these systems are coupled to others, and are not intrinsic to the model. For more details on symmetries and  gauge invariance in this context, see Appendices   \ref{sec:cons-line-quadr} and \ref{sec:gauge-invar-sympl}, respectively.

\section{Lumped-element electrical networks}
\label{Sec:QuantizationLumpedNetworks}
In order to derive a systematic, canonically quantized theory of superconducting lumped-element circuits, it is useful to take as an starting point a classical Lagrangian~\cite{Devoret:1997,Vool:2017}. The additivity of components in the Lagrangian is immediately obvious, while additivity for  the Hamiltonian is not necessarily as straightforwrd. Furthermore, a nonsingular Lagrangian determines  the Poisson brackets, and one does  not need to  inferr them  to correctly match Kirchhoff's equations of motion~\cite{Feynman:2010}. One of the problems in the analysis of circuitry, therefore, consists on the construction of nonsingular Lagrangians in a systematic way. This objective, however, has hitherto compelled  us to restrict the type of circuit to be analyzed so that there is a well defined mechanism for the identification of independent variables that still does provide us with a nonsingular Lagrangian.

In contrast, we propose to construct directly a nonsingular Lagrangian and eliminate redundancies by the systematic application of Williamson's symplectic diagonalization. Thus we consider a quite possibly redundant configuration space, such that all linear energy bearing lumped elements are assigned to a kinetic energy. In this way we are sure of the applicability of the Legendre transform. We shall see that there are two sources of redundancy, which, for the linear systems that we consider, correspond to the nondynamical and free particle degrees of freedom mentioned above.

This path followed here is inspired by the one taken to describe transmission lines  coupled by ideal NR elements in~\cite{ParraRodriguez:2022}, where a doubled configuration space description of the lines was introduced to  completely determine the normal mode structure of the combined system. As stated, the  kinetic energy will be the addition of the contributions of all lumped inductors and capacitors in the network. This will be the only set of linear energetic term of the lagrangian, which will be completed by linear nonenergetic constraint terms and relevant nonlinear energetic terms. Naturally a reduction of variables in the configuration space is useful to simplify the algorithm, but we are going to work on worst-case (brute force) scenarios. For the sake of completeness, we present the Lagrangian terms describing the nonreciprocal ideal elements  \cite{ParraRodriguez:2019,ParraRodriguez:2022,Rymarz:2018,Rymarz:2021}, the Belevitch transformer  \cite{Solgun:2015}, the nonlinear elements like the Josephson and the phase-slip junction \cite{Arutyunov:2008,Astafiev:2012} and discuss their effect on the Hamiltonian derivation.

\subsection{Linear reciprocal networks}
\begin{figure}[h]
	\includegraphics[width=.6\linewidth]{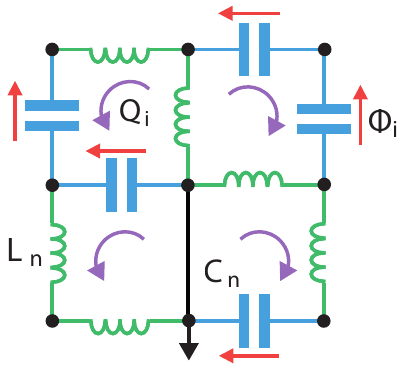}
	\caption{LC network. Asigning loop-charge (branch-flux) variables for the inductors (capacitors)  it is possible to write systematically a nonsingular kinetic Lagrangian term in a double space.}
	\label{fig:LC_network}
\end{figure}

Historically, the analysis of lumped element networks has been performed using the language of graph-theory~\cite{Devoret:1997,Burkard:2004,Burkard:2005}. Let us consider a common linear reciprocal network of 2-terminal lumped elements consisting on a set of inductors and capacitors connected in a {\it tree} containing $N$ number of nodes and $M$ number of loops such that $M\leq N$, as in the example in Fig. {\ref{fig:LC_network}}. 

Currents and voltages are denoted by time derivatives of loop charges $\dot{Q}\equiv \partial_t Q$ and node fluxes $\dot{\Phi}$ respectively. However, instead of using  node-flux variables, let us assign a branch-flux variable for every capacitor ($N_C$), and a loop-charge variable for every inductor ($M_L<M$). The equations of motion of such a generic network can be written in the form
\begin{align}
	\msC\ddot{\bPhi}+\msD^T\dot{\bQ}&=0,\label{eq:EOMs_C_D}\\
	\msL\ddot{\bQ}-\msD\dot{\bPhi}&=0.\label{eq:EOMs_L_D}
\end{align}
where $\msC$ and $\msL$ are positive diagonal matrices (therefore both full rank), and $\msD$ is an {\it adjacency} matrix encoding the topology of the terminal connections. This set of equations are a generalized discrete limit of the telegrapher's equations in a transmission line \cite{Pozar:2009,ParraRodriguez:2022} and flow from the following Lagrangian, written in the doubled (doubled as both fluxes and charges appear) configuration space description as
\begin{equation}
	L_{\mathrm{LC}}=\frac{1}{2}\dot{\bPhi}^T\msC\dot{\bPhi}+\frac{1}{2}\dot{\bQ}^T\msL\dot{\bQ}-\dot{\bQ}^T\msD\bPhi.\label{eq:Lagrangian_LC_ds}
\end{equation}
 The total apparent number of degrees of freedom is $N_C+M_L$, but there are  constraints given by  $\msD$ and $\msD^T$. In fact, it is  well known  that the number of node-flux and of loop-charge variables are  upper bounds to the number of real harmonic degrees of freedom of a network~\cite{Ulrich:2016}.  In contrast to previous works on canonical quantization of electric networks, let us assume for the sake of generality that we describe the Hamiltonian with redundancy, i.e., with a configuration space of  $N_C+M_L$ variables.

The invertibility of the kinetic matrices $\msC$ and $\msL$ permits  a Legendre transformation to obtain the Hamiltonian
\begin{equation}
	H_{\mathrm{LC}}=\frac{1}{2}\bPi^T\msC^{-1}\bPi+\frac{1}{2}\left(\bP-\msD\bPhi\right)^T\msL^{-1}\left(\bP-\msD\bPhi\right),\label{eq:H_LC_network}
\end{equation}
where the conjugate variables are $\bPi=\partial L/\partial \dot{\bPhi}$ and $\bP=\partial L/\partial \dot{\bQ}$. The dimension of the phase space is thus $2\times(N_C+M_L)$. This Hamiltonian is, on its face, positive semidefinite, as $\mathsf{C}^{-1}$ and $\mathsf{L}^{-1}$ are positive definite. The zero modes of the Hamiltonian will necessarily be associated with the possibility that $\bP-\msD\bPhi$ be zero. By Williamson's theorem, there exists a canonical transformation that expresses the Hamiltonian as describing a  set of free particles, harmonic oscillators, and nondynamical conjugate pairs of variables, see Appendix~\ref{app:LC_networks}. Studying the kernels of $\msJH$ and $(\msJH)^2$, it can be easily proven that number of nondynamical pairs is the same as that of harmonic oscillators $n_{nd}=n_{ho}=[N_C+M_L-n_f]/2$. The free particles are due to ignorable coordinates in the Hamiltonian, i.e. due to some translation symmetry. Those arise from the reducibility of the initial circuit by series/parallel equivalences. The nondynamical sector, on the other hand, turns out to be there because we have indeed doubled the number of variables, once the full reduction in terms of those equivalences has been carried out.

Let us illustrate the idea with a trivial example of a harmonic LC-oscillator with unnecessarily redundant descriptions.
\begin{figure}[h]
	\includegraphics[width=1\linewidth]{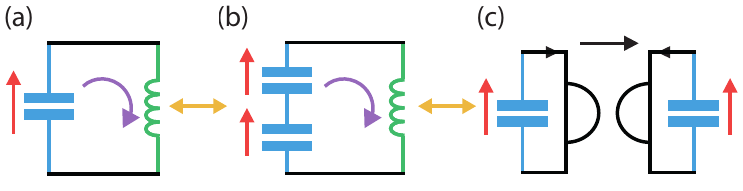}
	\caption{(a) Equivalent representations of a harmonic oscillator with: (a) an LC circuit; its minimal representation, (b) two capacitors in series and (c) a gyrator and a capacitor replacing the inductor. Redundant representations (b, c) may contain free-particle and/or nondynamical degrees of freedom.}
	\label{fig:LCC_circuit}
\end{figure}
The circuits in Fig. \ref{fig:LCC_circuit} are three different circuital representations of an LC harmonic oscillator. It is worth realizing that even its simplest version (a) admits irreducible representations in terms of only a flux $(\Phi)$ or charge coordinate $(Q)$, i.e., the equations of motion can be written in terms of a single variable, without requiring a redundant description with both variables in the configuration space. Even more, the circuit in (b) has two capacitors in series such that the new set of flux coordinates $(\Phi_1, \Phi_2)$ adds an extra level of redudancy. Indeed, upon the use of the sum rule for series capacitances $C_s=C_1\parallel C_2=(C_1+C_2)/C_1 C_2$, it is possible to eliminate a flux coordinate to reach (a). 

We now apply mechanically the method described above to circuit (b). That means assigning a variable and a kinetic energy term to each capacitor and inductor of the system, and computing the adjacency matrix to determine the constraint terms. Thus  Lagrangian (\ref{eq:Lagrangian_LC_ds}) should now be used, with matrices
\begin{align}
	\msC=\begin{pmatrix}
		C_1&0\\0&C_2
	\end{pmatrix},\quad \msL=\begin{pmatrix}
		L
	\end{pmatrix},\quad\msD=\begin{pmatrix}
		1&1
	\end{pmatrix},
\end{align}
with vector variabless $\bPhi=(\Phi_1,\Phi_2)^T$ and $\bQ=(Q)$. The Euler-Lagrange equations (\ref{eq:EOMs_C_D},\ref{eq:EOMs_L_D}) reveal the big redundancy in the description. Traditionally, one would eliminate the charge $Q$ by integrating Eq. (\ref{eq:EOMs_L_D}) in time while setting to zero the constant initial charge, i.e., $\dot{Q}=L^{-1}\msD{\bPhi}$, in which case, eq. (\ref{eq:EOMs_C_D}) simplifies to $\msC\ddot{\bPhi}=\msD^T\msL^{-1}\msD\bPhi$.
Given these equations, one realizes that a simpler Lagrangian exists, and, after
a  contact transformation of coordinates, $\bPsi=\msf{E}\bPhi$, the harmonic oscillator nature of the system is revealed with a new Lagrangian $L=\frac{1}{2}(C_s\dot{\Psi}^2-L^{-1}\Psi^2)$, where $C_s=(C_1+C_2)/C_1 C_2$. 

However, in our proposal we  do not need to perform at this stage the circuital reduction, and in fact it will come out the systematics. Namely, we write Hamiltonian (\ref{eq:H_LC_network}), and apply Williamson's theorem. That is, we find  a symplectic-orthonormal basis such the Hamiltonian matrix in this basis is diagonal. This computation reveals the presence of  a nondynamical variable and a free particle in addition to the harmonic oscillator. The free particle is directly related to the reducibility to an equivalent capacitor of the two series capacitors, while the nondynamical variable is a consequence of our redundant description. See Appendix \ref{app:LC_networks} for further details.

\subsection{Nonreciprocal linear ideal devices}
\begin{figure}[h]
	\includegraphics[width=.7\linewidth]{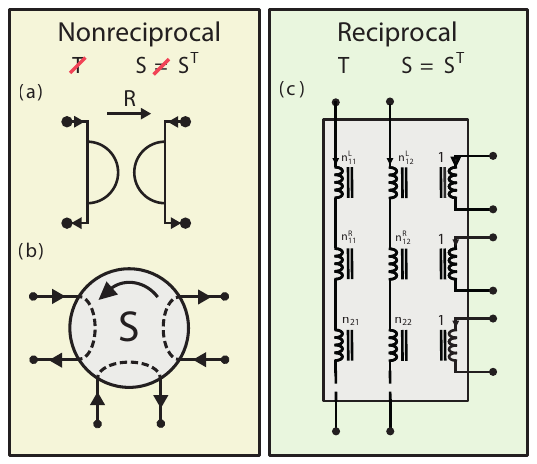}
	\caption{Multiport lossless linear system. The (a) gyrator and (b) circulator are 2- and 3-port ideal nonreciprocal elements implementing constraints between flux and charge variables and break of time-reversal symmetry ($\mcl{T}$). A (c) right-hand Belevitch transformer yields flux-flux and charge-charge constraints.} 
	\label{fig:Mport}
\end{figure}

Having understood the analysis for the case of  linear lumped energetic elements, let us now discuss the introduction of the nonreciprocal ideal ones, which do not store energy. In particular we address those  breaking  time-reversal symmetry. These multi-terminal devices represent \emph{constraints} between fluxes and charges, see Fig.~\ref{fig:Mport}(a,b) as expressed in their constitutive equation  
\begin{equation}\label{eq:Sconstitutive}
	(1-\msS)\dot{\bar{{\bPhi}}}=R(1+\msS)\dot{\bar{{\bQ}}},
\end{equation}
written in terms of a nonsymmetric scattering matrix $\msS$($\neq \msS^T$), which always exists~\cite{Carlin:1964}. The overbar for the fluxes and charges denotes that they have been rescaled dimensionally, for ease of notation in the constitutive equation. In fact, such elements are more easily introduced in the Lagrangian formalism in terms of  their impedance and/or admittance descriptions, namely 
\begin{eqnarray}
	L_{\msY}=\frac{1}{2}\dot{\bar{{\bPhi}}}^T\msY\bar{\bPhi},\quad \text{and}\quad L_{\msZ}=\frac{1}{2}\dot{\bar{{\bQ}}}^T\msZ\bar{\bQ}, \label{eq:NR_constitutive_YZ}
\end{eqnarray}
with ports described by branch-flux or loop-charge coordinates respectively. Observe that in these descriptions only the antisymmetric part of either admittance or impedance can have an effect on the equations of motion, since the symmetric part gives a total time derivative. Further, notice that there is no energy associated with these magnetic terms.

The general nonreciprocal constitutive equation described by $\msS$ is more directly read as a constraint on the evolution of the flux and charge variables, albeit nonholonomic. This nonholonomic character suggests that it might best be treated with tools other than Lagrangians on a flux coordinate space. However, Eq. \eqref{eq:Sconstitutive}  can always be framed in Lagrangian terms as one of Eqs. \eqref{eq:NR_constitutive_YZ} as long as one accepts the possibility of  immittance matrix being singular, in which case one further requires an additional flux or charge constraint (which can enter as another Lagrange multiplier) at the ports \cite{ParraRodriguez:2019,ParraRodriguez:2021}, i.e. 
\begin{eqnarray}
	\msP_{-}\dot{\bPhi}&=&0,\quad\text{if }-1\in\lambda_{\msS},\\
	\msP_{+}\dot{\bQ}&=&0, \quad\text{if }\quad\,1\in\lambda_{\msS}.
\end{eqnarray}
In either case, starting from the doubled configuration space construction,  a Legendre transformation will be available after this reduction is properly dealt with. We remark the equivalence of the coupling terms in (\ref{eq:NR_constitutive_YZ}) with the $\msD$-term in Lagrangian  (\ref{eq:Lagrangian_LC_ds}). In fact, there is a mapping between both expressions by rescaling some of the flux (charge) variables by $R^{-1}$ ($R$), i.e. $\tilde{Q}_i=\Phi_i/R$. This is the reason for the enlarged configuration space approach, involving both fluxes and charges, to be natural in dealing with circuits with nonreciprocal elements. In other words, one can think of an inductor (capacitor) in a circuit always as an equivalent capacitor (inductor) when  seen through a gyrator, see Fig.~\ref{fig:LC_network}(c), in which one of the capacitors can be moved (in the sense of equivalence of circuits) through the gyrator, becoming thus an inductor, providing us with a read on the circuit as an LC oscillator.


\subsection{Belevitch transformers}
The final ideal passive linear element to be considered is the Belevitch transformer \cite{Belevitch:1950}, another even (multi) $2N$-terminal device which represents direct constraints between voltages and currents on $N_L$ pairs of terminals on the left side, and $N_R$ pairs on the other ($N=N_L+N_R$), see Fig.~\ref{fig:Mport}(c). As such, they can be written as Lagrange multipliers, and can be directly eliminated before the Legendre transformation in the configuration, see further in \cite{Solgun:2015,ParraRodriguez:2018}. They can be classified in right (R), and left (L) transformers, such that their constraints are expressed as the Lagrangian terms
\begin{eqnarray}
	L_{\mathrm{ET}}=\msf{\Lambda}_{E}\dot{\bQ}+ \msf{\Xi}_{E}\dot{\bPhi},\label{eq:L_Belevitch}
\end{eqnarray}
where $E=\{R, L\}$, and
\begin{eqnarray}
	\msf{\Lambda}_{R}\dot{\bQ}&=&\msf{\Lambda}(\dot{\bQ}_R+\msN\dot{\bQ}_L),\\
	\msf{\Xi}_{R}\dot{\bPhi}&=&\msf{\Xi}(\dot{\bPhi}_L-\msN^T\dot{\bPhi}_R),
\end{eqnarray}
and analogously for $L_{\mathrm{LT}}$, where $\msN$ is the turn-ratios (rectangular) matrix describing the transformer. Such elements are known to be required for the systematic analysis and synthesis of generic linear electrical systems. For instance, in Eq. (\ref{eq:EOMs_L_D}) we have assumed the inductance matrix to be full rank. However, this may not be the case generally (say when tighly-coupled mutual inductances appear, e.g., $M_{ij}=\sqrt{L_i L_j}$). In such situations, the system could,  among other possibilities, be written in terms of transformers and a full rank inductance matrix. In our context, Belevitch transformers will not impinge on the nonsingular character of the Lagrangian, and insofar as they introduce constraints and therefore redundancies they are taken care of in the Williamson analysis.

\subsection{Nonlinear lumped elements}
\begin{figure}[h!]
	\includegraphics[width=.75\linewidth]{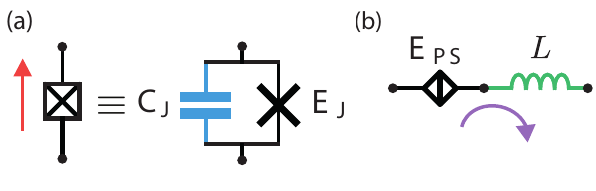}
	\caption{Conventional symbols for (a) a Josephson junction lumped element, with its intrinsic parallel capacitance, (b) and its dual circuit, a phase-slip junction with a series inductor.}
	\label{fig:EJ_EPS}
\end{figure}

Finally, let us discuss the addition of the nonlinear elements typically used in superconducting circuits, the Josephson junction (JJ)~\cite{Josephson:1962}, and its dual element, the {\it phase-slip} junction \cite{Arutyunov:2008,Astafiev:2012}. The unitary dynamics of the Josephson junction can be modeled by a linear capacitor in parallel with an element whose current-flux relation is nonlinear, $I_J=I_c \sin(2\pi\Phi_J/\Phi_0)=\partial_{\Phi_J}U(\Phi_J)$. The complementary phase-slip element also presents a sinusoidal nonlinear relation, in this case voltage-charge, $V_{PS}=V_c \sin(\pi Q_{PS} /2e)$, see Fig. \ref{fig:EJ_EPS}.
The nonlinear nature of these elements suggests their inclusion in a Lagrangian description in the form of  potential terms
\begin{eqnarray}
	L&=&L_{\mathrm{linear}}-U_J(\bPhi_J)-U_{PS}(\bQ_{PS}).
\end{eqnarray}
No issue arises in moving to a Hamiltonian description, as both flux and charge variables belong to the configuration space \emph{and} because a Josephson (phase-slip) junction will always be in parallel (series) to a capacitor (inductor), thus ensuring that a kinetic term always exists for the corresponding variable and mantaining the nonsingular character of the Lagrangian, 
\begin{equation}
	H=H_{\mathrm{linear}}+U_J(\bPhi_J)+U_{PS}(\bQ_{PS}).
\end{equation}

Once the complete classical and nonlinear Hamiltonian has been  obtained, several different strategies could be applied for its quantum analysis, and additional considerations come into play. For definiteness, let us  focus on circuits with Josephson junctions. It is now well established that the topology of the connection of the superconducting islands in a chip can dramatically change in which  way we are to model these nonlinear terms~\cite{Devoret:2021}. In the analysis of the Josephson effect one sees that a real junction island must be modeled by a compact phase variable $\varphi_J=\Phi_J\left(\frac{2\pi}{\Phi_0}\right)\in [0,2\pi]$, where $\Phi_J$ is the flux variable of the junction and  $\Phi_0$ the flux quantum constant. This has an impact on which canonical pair to consider in quantization, since the boundedness of the phase variable forces its conjugate operator to have a discrete spectrum, if selfadjointness is to be mantained. This is conveniently expressed with commutation relations of the form $[\hat{n}_J, e^{\pm i\varphi_J}] =\pm e^{\pm i\varphi_J}$ and the demand that $\hat{n}_J$ have a discrete (integer) spectrum. The experimental data confirming this idealization is in fact the discreteness of the spectrum of energies.

However, if the Josephson junction is  biased by an inductance, necessarily the phase variable decompactifies and the  spectrum of its quantization is  the whole real line $\varphi\in\mathbb{R}$. The conjugate operator is also a generator of translations in this case, but its spectrum is not discrete in this situation. The potential energy term is now a periodic function on the real line, instead of being defined on the circle. To see the difference in character, consider two free particle Hamiltonians of identical form, namely the usual kinetic term, in the two situations of compact and full real line variable. In the first case there is a discrete spectrum of eigenenergies, while for the real line all the eigenstates of energy are scattering states, and the spectrum is continuous.

The first question before us in this regard is whether our proposed analysis is applicable in either case, and the second one, if indeed applicable, is what the best strategy can be. The answer to the first question is that \emph{conceptually} it is only applicable to \emph{noncompact} variables. In the form we have presented the argument to this point  we have assumed systematically that the configuration space is indeed a linear space, in such a way that phase space is also a linear space and linear symplectic transformations are global and canonical. If we are to consider more general manifolds it is a fact that locally we can use Williamson's theorem to diagonalize symplectically the quadratic part of the Hamiltonian, but the global validity and usefulness of the construction remain to be explored, and we set that exploration aside for future work.

\begin{figure}[h!]
	\includegraphics[width=.8\linewidth]{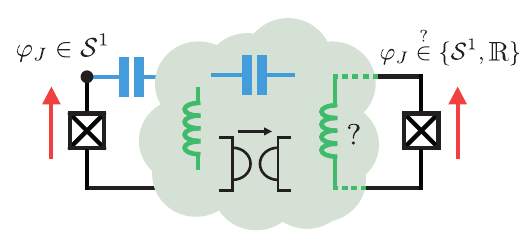}
	\caption{The Josephson junctions can always be modeled with compact variables in a general network if they appear in pure islands (left). In more general topological connections (right), it cannot be known {\it a priori}.}
	\label{fig:varphi_J_compact_or_real}
\end{figure}

Even within this limited purview, we shall apply the method in situations in which there are compact variables present.  Observe again that the presence of these compact variables in superconducting circuits is highly dependent on the topology of the circuit, see Fig.~\ref{fig:varphi_J_compact_or_real}. The central physical criterion, nonetheless, is that a proper superconducting island will be associated with a compact variable. In the class of circuits we consider this will be ensured if a Josephson junction is in series with a capacitor as part of a one port element.

Now assuming that there are indeed compact variables, we examine two situations in which we can apply the Williamson analysis. A first case takes place when the harmonic oscillator approximation is adequate for compact variables and, as long as the small oscillations condition is maintained, one can effectively consider the Josephson flux variable as decompactified. One such situation is that of the transmon qubit, with compact flux variable~\cite{Koch:2007}. The kinetic energy term is due to the Josephson capacity $C_J$, and its characteristic kinetic energy scale, $E_C$, behaves as  $E_C\propto\frac{1}{\tilde{C}_J}$. The potential energy due to the Josephson inductance presents a  characteristic potential energy scale  $E_J=I_c\left(\frac{\Phi_0}{2\pi}\right)$. Under the condition that $E_C\ll E_J$ the low energy sector can be studied in the  harmonic approximation, with additional small quartic terms (Duffing oscillator) $U(\Phi_J)\approx \frac{\Phi_J^2}{2 L_J} - U\frac{\Phi_J^2}{4!}$, with $L_J^{-1}=E_J\left(\frac{2\pi}{\Phi_0}\right)^2$ and $U=E_J\left(\frac{2\pi}{\Phi_0}\right)^4$. On coupling to a linear system, therefore, one can study the quadratic Hamiltonian, with inclusion of the quadratic part of the Josephson inductance, with the Williamson diagonalizing procedure. Once that has been carried out one should check that the weak nonharmonicity is mantained in the new diagonalizing phase space variables. 

An alternative way of stating that this possibility, namely of being able to apply the symplectic diagonalization procedure to the compact case, is that if the minima of the compact sector are very deep and essentially remain deep after the procedure then it can be relied upon.
Other than the issue of  compact/noncompact variables, the question remains as to whether the nonlinear variables are to be kept separate or, rather, taken into account in an application of the Williamson procedure. The criterion is the usefulness of either choice in providing us with tools for the quantum treatment of the system.

In the transmon regime mentioned above, for which the nonharmonicity is weak in the fashion presented above,  it would be sensible to perform Williamson's symplectic transformation on the whole set of coordinates in the network, i.e., including the flux variable of the Josephson junction in the analysis as a position variable and carrying out the harmonic approximation for its potential. Thus, in this regime, if we have a network with capacitors, inductors, NR elements, and JJs, one could start with a maximal set of coordinates $\bX=(\bPhi_J,\bPhi_C,\bQ_L)$, where $\bPhi_C$ are fluxes other than those  of the junctions, perform the harmonic approximation and symplectic diagonalization, and later check that the excursions of the coordinates inside the nonlinear terms are indeed small.

A second interesting case in which we can use the quantization procedure associated with Williamson diagonalization in the presence of compact variables is when the Lagrangian coupling of those compact fluxes to the rest of the circuit can be written as involving only their time derivatives. Observe however that in so doing one must ensure that the compact character is maintained properly. We will see a complete example of this idea in the next section. Then we can carry out a Williamson analysis of the Hamiltonian, in which the compact variables would appear as ignorable, i.e. as free particles, and after diagonalization add the periodic potentials. 

More generally, one could envisage carrying out the linear Williamson analysis just on the non JJ subsector of the phase-space, and only then consider the coupling to the Josephson junctions, in all their nonlinearity, to the rest of the system. In the next section we delve more deeply on the two possibilities, with the objective of expanding the methodology of blackbox quantization to generic immitance descriptions.

\section{Nonreciprocal black-box admittance quantization}
\label{Sec:JJs_Y_NR_BB}
As of late there has been an increasing interest in the use of black-box linear models to describe complex 3D volumes in superconducting circuitry~\cite{Nigg:2012,Solgun:2014,Solgun:2015,Solgun:2019,Minev:2020,Solgun:2021}. Most of them rely on having an immitance description of the linear system, and  typically  an impedance decomposition is used. This bias towards one of the descriptions is linked to (i) the generally prevalent preference of flux coordinates as configuration-space independent variables for the whole network, which is in turn due to the interest in Josephson junctions, for which it is most convenient as the nonlinearity is expressed as a potential energy, and (ii) the fact that the derivation of a Hamiltonian for the admittance expansion is more easily performed in terms of loop-charge variables. 

In this section we use the systematic Williamson diagonalization, as applied to an enlarged configuration space, to quantize in a consistent and systematic way a two-port nonreciprocal black-box generic admittance $\msY(\omega)$ capacitively coupled at its ports to Josephson junctions. We  show that it does not present any further obstacle with respect to its impedance courterpart. In particular, if one were to use a purely flux-variable based quantization, the Lagrangian of this circuit would be necessarily written in terms of a {\it singular} kinetic (capacitance) matrix. We sidestep this singularity by working in an enlarged configuration space~\cite{Ulrich:2016}.

\begin{figure}[h]
	\includegraphics[width=1\linewidth]{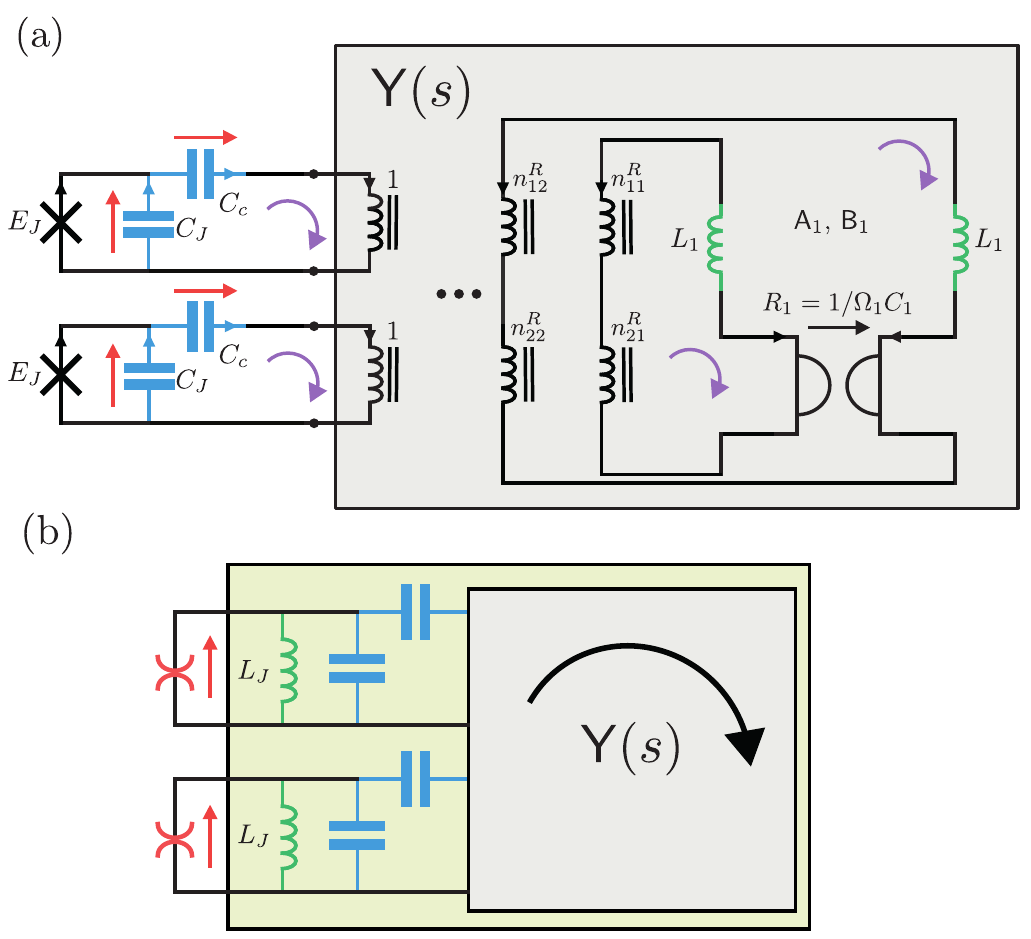}
	\caption{Josephson junctions capacitively coupled to a two-port nonreciprocal device $\msY(\omega)$ explicitly showing the (a) multiport Foster representation of the admittance. (b) The Josephson junctions are broken into a quadratic ($L_J$) and quartic and higher potential terms.}
	\label{fig:2CQ_2_port_NR_Y}
\end{figure}

The general Foster decomposition of an admittance response is composed of inductor-gyrator series stages connected to the output ports through a Belevitch transformer, see Fig. \ref{fig:2CQ_2_port_NR_Y}. For the sake of simplicity, and without loss of generality, we restrict the analysis to a 1-stage resonance circuit. From a physical perspective, the Belevitch transformer finally yields {\it oriented} energy participation ratios between the internal (non)reciprocal modes in the black-box and the external ports~\cite{Minev:2020}.

In line with the discussion above, we quantize along two different routes: (i) by expanding the cosine potentials of the JJs and diagonalizing the quandratic part of the Hamiltonian of the whole system, and (ii) by diagonalizing a harmonic subsector having set aside the JJs coordinates for later (which would permit the inclusion of the whole cosine in the posterior analysis). We show the match of both paradigms in the full linear approximation, i.e., two-step versus a one-step symplectic diagonalizations.
 
The  Lagrangian for Josephson junctions coupled to a generic admittance can be written by  adding up energetic and constraint terms for all the different lumped elements, i.e., $L=L_{LC}+L_{J}+L_{LT}+L_{\msY}$. Here $L_{LC}$, $L_{\msY}$ and $L_{LT}$ are defined in Eqs. \eqref{eq:Lagrangian_LC_ds}, \eqref{eq:NR_constitutive_YZ} and \eqref{eq:L_Belevitch}, respectively, and we collect the JJ terms in $L_J$. By elimination of the trivial constraints, arising from the Belevitch right fluxes and left charges through the Lagrange multipliers (\ref{eq:L_Belevitch}), it can be rewritten as 
\begin{align}\label{eq:L_blackbox}
	L=\,&\frac{1}{2}\left(\dot{\bPhi}^T\msC\dot{\bPhi}^T+\dot{\bQ}^T\msL\dot{\bQ}-2\dot{\bQ}^T\msD\bPhi+\dot{\bQ}^T\msZ\bQ\right)\nonumber\\
	&-U(\bPhi_J),
\end{align}
with  flux coordinates  describing the external variables $\bPhi^T=(\bPhi_J^T,\bPhi_c^T)=(\Phi_{Ja},\Phi_{Jb},\Phi_{ca},\Phi_{cb})$, and charge coordinates for the internal variables in the admittance representation $\bQ^T=(Q_{L1},Q_{L2})$.  The matrices read for the concrete example of Fig. \ref{fig:2CQ_2_port_NR_Y} (a) 
\begin{align}\label{eq:example}
	\msC&=\begin{pmatrix}
		\msC_J&0\\0&\msC_c
	\end{pmatrix},\quad\msL=L_1\mone_2,\quad \msD=\msD_R\msN^T\msD_L\\
	\msD_L&=\begin{pmatrix}
		1&0&1&0\\0&1&0&1
	\end{pmatrix}=-\msD_R,\quad
	\msN=\begin{pmatrix}
		n_{11}&n_{12}&0&0\\0&0&n_{21}&n_{22}
	\end{pmatrix},\nonumber
\end{align}
where $	\msC_J=C_J\mone_2$, $\msC_c=C_c\mone_2$, and $\msZ=-R\msJ_2$. On the other hand, the nonlinear potential of the modeling nonlinear Josephson element is $U(\bPhi_J)=-\sum_i E_{Ji}\cos(\varphi_{Ji})$. A Legendre transformation is readily available and results in the Hamiltonian 
\begin{align}\label{eq:H_JJs_Cc_Ymat}
	H=&\,\frac{1}{2}(\bP-\tfrac{1}{2}\msZ\bQ+\msD\bPhi)^T\msL^{-1}(\bP-\tfrac{1}{2}\msZ\bQ+\msD\bPhi)\nonumber\\
	&+\frac{1}{2}{\bPi}^T\msC^{-1}{\bPi}^T +U(\bPhi_J)\,,
\end{align}
where the conjugate canonical pairs are defined by the Poisson brackets  $\{\bPhi,\bPi^T\}=\{\bQ,\bP^T\}=\mone$. Please, notice the similarity between the coupling of branch-fluxes $\bPhi$ and conjugated flux momenta $\bP$ through the skew-symmetric matrix $\msZ$ and the minimal-coupling Hamiltonian in the Landau problem. 

Having obtained this Hamiltonian, it is a pedagogical exercise to perform an analytical study of e Hamiltonian (\ref{eq:H_JJs_Cc_Ymat}) in the absence of the Josephson junctions, i.e. setting $U$ to be zero. Given that $\mathrm{dim}\left(\mathrm{ker}\left[\msD^T\right]\right)=0$, which happens whenever the transformer matrix is nontrivial, the kernels of $K_1$ and $K_2$ can be parametrized as 
\begin{align}
	\bv_1=\begin{pmatrix}
		\bq\\ \bsb{\phi} \\ \frac{\msZ}{2}\bq-\msD\bsb{\phi}\\0
	\end{pmatrix}\in K_1,\,\,
	\bv_2=\begin{pmatrix}
	\bq\\ \bsb{\phi} \\ \frac{\msZ}{2}\bq-\msD\bsb{\phi}\\ \msC \be_{\msD}
\end{pmatrix}\in K_2\,.
\end{align}
Here $\be_{\msD}$ is a generic element of $\mathrm{ker}\left[\mathsf{D}\right]$, and  $\bsb{\phi}$ and $\bq$ span vector spaces of dimension $N_C$ and $M_L$, respectively. Note that the coordinates are written in the order $(\bQ^T,\bPhi^T,\bP^T,\bPi^T)$. The dimensionality can be directly read as  $\mathrm{dim}(K_1)=N_C+M_L$, and  $\mathrm{dim}(K_2)=N_C+M_L+\mathrm{dim}\left(\mathrm{ker}\left[\msD\right]\right)$, such that the number of free particles is $n_f=\mathrm{dim}\left(\be_{\msD}\right)=2$, and $n_{nd}=\frac{1}{2}(N_C+M_L-n_f)=n_{ho}=2$. It can be easily proven that removing a pair of capacitors, e.g.,  $C_{ci}$, $\mathrm{dim}\left(\mathrm{ker}[\msD]\right)=\mathrm{dim}\left(\mathrm{ker}\left[\msD^T\right]\right)=0$, and no free particles would appear in this description with such a choice of coordinates. In other words, by taking together the limits $C_{ci}\rightarrow\infty$, $\bPhi_c\rightarrow\bPhi_J$ and $\bPi_c\rightarrow\bPi_J$, the associated free-particle dynamics vanish, i.e., in algebraic terms $\mathrm{ker} [\msJH]=\mathrm{ker}[(\msJH)^2]$.

However, in presence of the Josephson junctions the corresponding fluxes are compact variables, as signalled above, and one should be more cautious in the application of the symplectic diagonalization idea. Observe to this point that the coupling of the Josephson fluxes to the rest of the system is through the term $\dot{\bQ}^T\msD\bPhi$ in Lagrangian \eqref{eq:L_blackbox}, which gives rise to the corresponding couplings in the Hamiltonian \eqref{eq:H_JJs_Cc_Ymat}. It is therefore convenient to perform a change of variables in configuration space, shifting the coupling to the time derivative of the Josephson fluxes, given by
\begin{align}
  \label{eq:qshift}
  \bQ = \tilde{\bQ}+ 2\msZ^{-1} \msD_J\tilde{\bPhi}_J\,.
\end{align}
This is accompanied by no change in the flux coordinates, $\bPhi_J=\tilde{\bPhi}_J$ and $\bPhi_c=\tilde{\bPhi}_c$. 
Observe that this point transformation does not change the topology of configuration space, and it induces a canonical transformation on the full phase space (whose topology remains unchanged as well), even taking into account that it is compact along some directions. The change of coordinates in momentum space reads
\begin{align}
  \label{eq:tildefinal}
  \bP &= \tilde{\bP}\,,\nonumber\\
  \bPi_c &= \tilde{\bPi}_c\,,\\
  \bPi_J &= \tilde{\bPi}_J+2 \msD_J^T\msZ^{-1}\tilde{\bP}\,.\nonumber
\end{align}
Under this canonical transformation, the new Josephson flux variables only appear in the potential energy terms, $U\left(\tilde{\bPhi}_J\right)$, and the couplings of the junctions to the rest of the system  appear only in the kinetic part of the Hamiltonian.


Let us now apply to this situation the two most common points of view in the literature regarding quantization of Josephson junctions connected to an impedance (as opposed to an admittance as we do here). Namely, either (i) a two tier process, in which we select the part of the Hamiltonian in which neither  the flux variables for the  JJs \emph{nor} their conjugate variables appear, symplectically diagonalize that part, and then study the full Hamiltonian with these new variables, or (ii) separating the cosine potentials in a quadratic and higher order terms, setting aside the quartic and higher terms, computing the normal modes of the quadratic part of the Hamiltonian (with the quadratic part of the cosines included), and then expanding the quartic and higher terms in terms of the new collective normal mode coordinates. This last approach is known for impedance coupling as the black-box approach~\cite{Nigg:2012,Solgun:2014,Solgun:2015}, see Fig.~\ref{fig:2CQ_2_port_NR_Y}(b). We complete the study by showing the equivalence of  both perspectives under some specific conditions.

To facilitate further analysis of this Hamiltonian, we  perform trivial rescaling (symplectic) transformations in such a way that all the new coordinates of phase space share the same dimensions. The inductive loop-charge variables (and their conjugated flux momenta) transform into $\bar{\bQ}=\sqrt{R}\tilde{\bQ}$ ($\bar{\bp}=\tilde{\bp}/\sqrt{R}$), whereas the capacitive branch-fluxes (and conjugated charge momenta) turn into  $\bar{\bPhi}_{c/J}=\msC_{c/J}^{-\frac{1}{4}}\msL^{\frac{1}{4}}\tilde{\bPhi}_{c/J}$  ($\bar{\bPi}_{c/J}=\msC_{c/J}^{\frac{1}{4}}\msL^{-\frac{1}{4}}\tilde{\bPi}_{c/J}$). The Hamiltonian with exclusion of the Josephson potentials is then parameterised in the example of Fig. \ref{fig:2CQ_2_port_NR_Y} (a), for these homogeneous dimension coordinates, in terms of the frequencies $\Omega_{J}=1/\sqrt{C_J L}$, $\Omega_c=1/\sqrt{C_c L}$, and $\Omega=R/L$. For numerical illustration (see Appendix \ref{App:JJs_Cc_Ymat})  we set the transformer turn-ratios matrix to $n_{11}=n_{12}=n_{22}=1$ and $n_{21}=0$, and we work with homogeneous frequencies $\Omega=\Omega_c=\Omega_J=1$.

Let us now show explicitly the two tier quantization of this system. Performing a symplectic transformation that only mixes the subset of coordinates $\bar{\bY}=(\bar{\bQ},\bar{\bPhi}_c,\bar{\bP},\bar{\bPi}_c)$, and diagonalizes $\msH_{\mathrm{lin}}$, i.e., $(\msS_{\bY}^{-1}\bY)^T=(\bar{\by}_{\mathrm{nd}}^T,\bar{\bpi}_{\mathrm{nd}}^T,\bar{\by}_{\mathrm{ho}}^T,\bar{\bpi}_{\mathrm{ho}}^T)$,we obtain (two) harmonic oscillators linearly coupled (through momenta) to the (dressed) nonlinear system 
\begin{align}
	H=&\sum_{n\in{ho}}^2\frac{\Omega_n}{2}(\bar{y}_n^2+\bar{\pi}_n^2)+\left(\bar{\by}^T\msf{G}+\bar{\bpi}^T\msf{M}\right)\bar{\bPi}_J\nonumber\\
	&+\sum_{n=1}^2\left[\Omega_{J}\bar{\Pi}_{Jn}^2-E_{Jn}\cos(\bar{\varphi}_{Jn})\right],\label{eq:H_JJs_Cc_Ymat_separated}
\end{align}
where $\{\bar{y}_i,\bar{\pi}_j\}=\delta_{ij}$, and $\msf{G}$ and $\msf{M}$ are coupling matrices between the Josephson momenta and the partially-diagonalized coordinates. A canonical quantization of this system can now be performed given that the nondynamical sector has been singled out and set to zero~\cite{ParraRodriguez:2021}, promoting to quantum operators only the dynamical ones. After reexpressing the harmonic conjugate variables  in terms of  annihilation and creation operators, we obtain the Hamiltonian 
\begin{align}
	\hat{H}/\hbar=&\sum_{n,m}^2\left[\Omega_n a_n^\dag a_n + (g_{nm}a_n+g_{nm}^*a_n^\dag)\bar{\Pi}_{Jm}\right]+H_{J},\nonumber
\end{align}
where $H_J$ stands for all the terms involving only the Josephson sector in (\ref{eq:H_JJs_Cc_Ymat_separated}), namely, the second line. Specific values for eigenfrequencies and coupling vectors for this set of parameters can be seen in Appendix~\ref{App:JJs_Cc_Ymat}. 

If on the other hand a black-box approach~\cite{Nigg:2012} is preferred, we  separate out the quartic and higher terms of the Taylor  expansion of the cosine potentials  of the junctions. The remaining quadratic part (in which  a $\bar{\bPhi}_J^T \msL_J^{-1}\bar{\bPhi}_J$ term appears) is susceptible to symplectic diagonalization. Observe however that the compactness of the Josephson fluxes has been discarded, and only in a transmonlike regime would this process be valid. After this symplectic diagonalization, quantize the resulting (four) harmonic oscillators, and expand the nonlinear potential in terms of normal modes to obtain the Hamiltonian
\begin{align}
	\hat{H}_{bb}=\sum_{n}^4\hbar\bar{\Omega}_n b_n^\dag b_n+E_{J}\sum_{l={1,2}}\sum_{m=2}^\infty\frac{(-1)^{m+1}\bar{\varphi}_{Jl}^{2m}}{(2m)!}.\label{eq:H_JJs_Cc_Ymat_bb}
\end{align}
It is worth remarking that the flux variables of the Josephson junctions depend in this case only on the dynamical sector after the appropiate symplectic transformations, i.e., $\bar{\varphi}_{Ji}\equiv\bar{\varphi}_{Ji}(\bz_{ho},\bpi_{ho})$, with $\bZ=\msS^{-1}\bar{\bX}=(\bz_{nd},\bz_{ho},\bpi_{nd},\bpi_{ho})$ thus setting the correct grounds for a low-lying energy investigation. Naturally, we remind the reader that the previously described routines match, as it can be easily proven by expanding to second order the cosine in Eq. (\ref{eq:H_JJs_Cc_Ymat_separated}) and performing a secondary symplectic transformation, see Appendix~\ref{App:JJs_Cc_Ymat} for details.

As a final note, we observe that in more general scenarios  the Josephson variables might indeed depend on the nondynamical coordinates of the linear problem, i.e., $ \bar{\varphi}_{Ji}(\bz_{nd}^T,\bpi_{nd}^T,\bz_{ho}^T,\bpi_{ho}^T)$. Thus the black-box approach as presented here  would not set out an immediately obvious path  for the investigation of the low-lying energy sector, and  the two tier process, e.g., Eq. (\ref{eq:H_JJs_Cc_Ymat_separated}), should always be preferred. Furthermore, the problem here stated would indeed increase in complexity if the Hamiltonian had extra time-dependent classical forces, as it is the case when dealing with pulses for quantum computation or readout~\cite{Malekakhlagh:2020,Petrescu:2020,Petrescu:2021}, and the two tier picture is recommended as the basis for further analysis.

\section{Conclusions and outlook}
\label{Sec:Conclusions}
In this article we have put forward a consistent quantization procedure for  superconducting lumped-element networks modeled in an enlarged configuration space description, with both flux and charge variables. The manifest redundancy is eliminated in the Hamiltonian through a symplectic transformation provided systematically by Williamson's theorem, for which we have given a constructive algorithm for real positive-semidefinite Hamiltonian matrices. To the best of our knowledge, this is its first explicit application to a simple physical scenario. The full-fledged Williamson's theorem has been recurrently referenced in the quantum physics community, but  only positive-definite matrices have been explicitly provided with either proof and explicit construction~\cite{Idel:2017,Ding:2021,de2006symplectic}. Given its symplicity, this algorithm will find use in a plethora of gaussian systems beyond electrical circuits.

In the context of the specific problem here tackled, we emphasize that the enlarged configuration space idea becomes a useful starting point for the identification and quantization of the dynamical degrees of freedom in circuits with frequency-dependent  nonreciprocal devices where a mixed charge-flux configuration space variables may result inevitable. This reinforces the idea we put forward for circuits with transmission lines coupled through ideal nonreciprocal devices~\cite{ParraRodriguez:2021}. Naturally, our algorithm will also prove useful to single out free particle dynamics in reciprocal circuits where a redundant choice of either flux or charge variables is chosen as the configuration space.

As an example of the power of our technique, we have reduced the problem of quantizing a circuit composed of Josephson junctions capacitively coupled to a general linear system described in terms of an admittance matrix to that of correctly identifying the dynamical and nondynamical sectors of a redundant Hamiltonian. Such analysis would have been much more cumbersome had we started the analysis with a Lagrangian written in terms of just flux variables, because the kinetic matrix would be singular. In fact, we have presented and compared the two most standard methods to  quantize canonically superconducting circuits, i.e., performing: (i) a partial diagonalization of the purely linear sector, setting aside at that point of the analysis the full nonlinear cosine potentials of the Josephson junctions, and (ii) a full diagonalization of the linear sector within a generalized black-box quantization paradigm. We point out that in some cases the black-box approach may require further refinement when the nonlinear potentials are  written in the transformed basis. This would arise from a  (possibly) nonlinear coupling between those sectors that in the linear approximation had been identified as dynamical and nondynamical, respectively.

Looking to the future, further work will be required to extend these ideas for an exact quantization prescription in the case of quasi-lumped circuits containing frequency-dependent nonreciprocal black-boxes, as well as  transmission lines and nonlinear elements, along the lines first suggested in~\cite{ParraRodriguezPhD:2021}.

\begin{acknowledgements}
I. L. E. acknowledges support of the Basque Government grant IT986-16. A. P.-R. thanks Basque Government Ph.D. grant PRE-2016-1-0284, and the Canada First Research Excellence
Fund. We thank C. Lled\'o-Veloso for a first proofreading the article. 
\end{acknowledgements}

\appendix
\section{Symplectic diagonalization of positive semidefinite Hamiltonians}
\label{sec:sympl-diag-posit}
As introduced in the text, we investigate the symplectic diagonalization of (even dimensional) positive semidefinite matrices. We shall denote them as $\mathsf{H}$, and the canonical symplectic matrix $\mathsf{J}$ is
\begin{equation}
	\label{eq:jmat}
	\mathsf{J}=
	\begin{pmatrix}
		0&\mone\\-\mone&0
	\end{pmatrix}\,.
\end{equation}
If necessary we will distinguish canonical symplectic matrices for different dimensions with explicit mention of the dimension.

Williamson's theorem concerns the canonical forms under symplectic transformations of even dimensional matrices over fields of non zero characteristic.
The crucial fact underlying the full Williamson's theorem and its restrictions is that the conjugacy normal forms of $\mathsf{JH}$ determine completely the symplectic normal forms of $\mathsf{H}$. This comes about because the space of states is a direct sum of symplectically orthogonal subspaces, and those are associated with generalised eigenspaces of $\mathsf{JH}$.

The detailed proof of Williamson's theorem for the real field consists in the analysis of each possible structure of generalised eigenspaces and the determination of the corresponding canonical form for that sector. We refer to \cite{Hoermander:1995} for the complete proof. Here we simply give a general guide to the concepts involved in the next subsection, and later on  we will concentrate on the case of interest, that of positive semidefinite Hamiltonians, taking the theorem as a starting point.

\subsection{Generalities of Williamson's theorem for real symmetric matrices}
\label{sec:gener-will-theor}
For real symmetric matrices, as is the case of interest to us, the characteristic polynomial of $\mathsf{JH}$, $p(\lambda)$, has real coefficients and is even, $p(-\lambda)=p(\lambda)$. This simple fact comes about because $\mathsf{J}^2=-\mone$ and $\mathrm{det}(\mathsf{J})=1$. Thus,
\begin{align}
	\label{eq:peven}
	p(\lambda)&= \mathrm{det}\left[\mathsf{JH}-\lambda\mone\right] = \mathrm{det}\left[\mathsf{H}+\lambda\mathsf{J}\right]\nonumber\\
	&= \mathrm{det}\left[\left(\mathsf{H}+\lambda\mathsf{J}\right)^T\right]= \mathrm{det}\left[\mathsf{H}-\lambda\mathsf{J}\right]\nonumber\\
	&= p(-\lambda)\,.
\end{align}
Therefore, if $\lambda$ is an eigenvalue (possibly generalised), then so are $-\lambda$,  the complex conjugate $\lambda^*$ and $-\lambda^*$. Thus there are in principle four types of eigenvalue: i) the generic $\mu+i \nu$ with real $\mu$ and $\nu$, in a quartet $\left\{\mu+i\nu,\mu-i\nu,-\mu-i\nu,-\mu+i\nu\right\}$; ii) purely imaginary $\pm i\mu$; iii) purely real $\pm \mu$; and iv) 0.

We shall use the notation $\langle\bullet,\bullet\rangle$ for the canonical inner product both for the real space $\mathbb{R}^{2n}$ on which $\mathsf{H}$ and $\mathsf{J}$ act and for the sesquilinear inner product of its complexification and of $\mathbb{C}^{2n}$. In matrix terms, $\langle\mathbf{u},\mathbf{v}\rangle= \mathbf{u}^\dag\mathbf{v}$, with $\dag$ denoting hermitian conjugate.

Let us now consider $\mathbf{v}_1$ and $\mathbf{v}_2$ proper eigenvectors of $\mathsf{JH}$, with eigenvalues $\lambda_1$ and $\lambda_2$ respectively. Then
\begin{align}
	\label{eq:symporth}
	\left(\lambda_2+\lambda_1^*\right)\langle\mathbf{v}_1,\mathsf{J}\mathbf{v}_2\rangle &= \left\langle\mathbf{v}_1,\mathsf{J}\left(\mathsf{JH}\mathbf{v}_2\right)\right\rangle+ \left\langle\mathsf{JH}\mathbf{v}_1,\mathsf{J}\mathbf{v}_2,\right\rangle\nonumber\\
	&= -\left\langle\mathbf{v}_1,\mathsf{H}\mathbf{v}_2\right\rangle+\left\langle\mathsf{H}\mathbf{v}_1,\mathbf{v}_2\right\rangle =0\,.
\end{align}
Therefore, if $\lambda_1^*+\lambda_2\neq0$ the corresponding two eigenvectors are symplectic orthogonal. In fact, we can extend this property to the generalised eigenspaces if there are nontrivial Jordan chains, by induction. To this point, let us consider that the respective Jordan chains are of the form
\begin{equation}
	\label{eq:jordanchain}
	\mathsf{JH}\mathbf{v}_i^{(k)}=\lambda_i\mathbf{v}_i^{(k)}+\mathbf{v}_i^{(k-1)}\,.
\end{equation}
Then, along the lines of Eq. \eqref{eq:symporth}, we have
\begin{align}
	\label{eq:symporthgen}
	&\left(\lambda_2+\lambda_1^*\right)\langle\mathbf{v}_1^{(k)},\mathsf{J}\mathbf{v}^{(l)}_2\rangle = \left\langle\mathbf{v}^{(k)}_1,\mathsf{J}\left(\mathsf{JH}\mathbf{v}^{(l)}_2-\mathbf{v}_2^{(l-1)}\right)\right\rangle\nonumber\\
	& + \left\langle\left(\mathsf{JH}\mathbf{v}^{(k)}_1-\mathbf{v}^{(k-1)}_1\right),\mathsf{J}\mathbf{v}^{(l)}_2,\right\rangle\nonumber\\
	&= -\left\langle\mathbf{v}^{(k)}_1,\mathsf{H}\mathbf{v}^{(l)}_2\right\rangle+\left\langle\mathsf{H}\mathbf{v}^{(k)}_1,\mathbf{v}^{(l)}_2\right\rangle \nonumber\\
	& -\left\langle \mathbf{v}_1^{(k)},\mathsf{J}\mathbf{v}_2^{(l-1)}\right\rangle-\left\langle \mathbf{v}_1^{(k-1)},\mathsf{J}\mathbf{v}_2^{(l)}\right\rangle\nonumber\\
	&= -\left\langle \mathbf{v}_1^{(k)},\mathsf{J}\mathbf{v}_2^{(l-1)}\right\rangle-\left\langle \mathbf{v}_1^{(k-1)},\mathsf{J}\mathbf{v}_2^{(l)}\right\rangle\,,
\end{align}
and, by induction, the respective chains are symplectic orthogonal if $\lambda_2+\lambda_1^*\neq0$.

Please observe that 
$\langle\bullet, \mathsf{J}\bullet\rangle$ is \emph{not} antisymmetric for complex vectors, namely, given two real vectors $\mathbf{u}$ and $\mathbf{v}$ one has
\begin{equation}
	\label{eq:hermandsymp}
	\left\langle\mathbf{u}+i\mathbf{v},\mathsf{J}\left(\mathbf{u}+i\mathbf{v}\right)\right\rangle = 2i\left\langle\mathbf{u},\mathsf{J}\mathbf{v}\right\rangle\,.
\end{equation}

Let $V_\lambda$ be the generalised eigenspace of $\mathsf{JH}$ for the eigenvalue $\lambda$. Then the result above is that the full complexified vector space $V=\mathbb{R}^{2n}_{\mathbb{C}}$ can be written as a direct sum of symplectic orthogonal subspaces,
\begin{equation}
	\label{eq:Vdecompose}
	V= V_0\oplus \bigoplus_{\lambda\neq0}\left(V_\lambda\oplus V_{-\lambda^*}\right)\,,
\end{equation}
where the sum avoids double counting, i.e. if $\lambda$ is in the index set, then $-\lambda^*$ is not. 

The relevant issue at this point is the construction of a canonical symplectic basis using these generalised eigenspaces. We remind the reader that a symplectic basis of $\left\{\mathbb{R}^{2n},\mathsf{J}\right\}$ is a basis with elements $\left\{\mathbf{e}_1,\mathbf{e}_2,\ldots,\mathbf{e}_n,\mathbf{f}_1,\ldots,\mathbf{f}_n\right\}$ such that
\begin{align}
	\label{eq:sympbasis}
	\left\langle \mathbf{e}_i,\mathsf{J}\mathbf{e}_j\right\rangle&=0\,,\nonumber\\
	\left\langle \mathbf{f}_i,\mathsf{J}\mathbf{f}_j\right\rangle&=0\,,\\
	\left\langle \mathbf{e}_i,\mathsf{J}\mathbf{f}_j\right\rangle&=\delta_{ij}\,.\nonumber
\end{align}

As $\mathsf{JH}$ is real, the generalised eigenvector $\mathbf{v}$ corresponding to a complex eigenvalue  $\lambda$ has an associated generalised eigenvector corresponding to  the complex conjugate eigenvalue $\lambda^*$, namely $\mathbf{v}^*$. Thus, in the construction of the \emph{real} symplectic basis we desire we shall have to use combinations of $V_\lambda$ and $V_{\lambda^*}$. Summarizing, there are four types of symplectic subspaces to consider. First, the generic $\lambda=\mu+i\nu$ with $\mu,\nu>0$, for which the symplectic space to be considered is the direct sum
\begin{equation}
	\label{eq:fourcomponent}
	V_\lambda\oplus V_{-\lambda^*}\oplus V_{\lambda^*}\oplus V_{-\lambda}\,.
\end{equation}
Secondly, the purely imaginary case for which we have the symplectic space $V_{i\mu}\oplus V_{-i\mu}$. Thirdly, the purely real case $V_\mu\oplus V_{-\mu}$, and finally $V_0$. As stated above, the theorem relies on the detailed analysis of these four cases separately. 

In the construction of the relevant symplectic basis it is important to notice that the generalised eigenspaces of the different elements of the quartet (or doublet) will have the same structure. That is to say, if $\lambda$ is a generalised eigenvalue with a Jordan chain of length $n_\lambda$, then $-\lambda$, $\lambda^*$ and $-\lambda^*$ will all have Jordan chains of the same length, and similarly for the doublets. Similarly, if $\lambda$ is degenerate then the degeneracy structure is shared by the other elements of the multiplet.

For definiteness, consider now a nondegenerate quartet, with shared Jordan chain length $n_\lambda$. We have a set of generalised eigenvectors of $\mathsf{JH}$, obeying the Jordan chain structure of Eq. \eqref{eq:jordanchain}, that we will denote as $\left\{\mathbf{v}^{(k)},\mathbf{u}^{(k)},\bar{\mathbf{v}}^{(k)},\bar{\mathbf{u}}^{(k)}\right\}$, corresponding to $\lambda,-\lambda^*,\lambda^*,-\lambda$ respectively. We use here $\bar{\mathbf{u}}$ to denote the complex conjugate in order to avoid an overburden of superscripts.
According to the general result above, $\left\langle \mathbf{v}^{(k)},\mathsf{J}\mathbf{u}^{(k)}\right\rangle$ is not trivial, but $\left\langle \mathbf{v}^{(k)},\mathsf{J}\bar{\mathbf{u}}^{(k)}\right\rangle=\left\langle \mathbf{v}^{(k)},\mathsf{J}\bar{\mathbf{v}}^{(k)}\right\rangle=\left\langle \mathbf{u}^{(k)},\mathsf{J}\bar{\mathbf{u}}^{(k)}\right\rangle=0$. Let us choose to normalize $\mathbf{v}^{(k)}$ and $\mathbf{u}^{(k)}$, including phases, such that
\begin{equation}
	\label{eq:normchoice}
	\left\langle \mathbf{v}^{(k)},\mathsf{J}\mathbf{u}^{(k)}\right\rangle=2\,.
\end{equation}
We now define the real vectors
\begin{align}
	\label{eq:aketc}
	\mathbf{e}_{2k-1}&=\frac{1}{2}\left[\mathbf{v}^{(k)}+\bar{\mathbf{v}}^{(k)}\right]\,,\nonumber\\
	\mathbf{e}_{2k}&=\frac{-i}{2}\left[\mathbf{v}^{(k)}-\bar{\mathbf{v}}^{(k)}\right]\,,\nonumber\\
	\mathbf{f}_{2k-1}&=\frac{1}{2}\left[\mathbf{u}^{(k)}+\bar{\mathbf{u}}^{(k)}\right]\,,\\
	\mathbf{f}_{2k}&=\frac{-i}{2}\left[\mathbf{u}^{(k)}-\bar{\mathbf{u}}^{(k)}\right],.\nonumber
\end{align}
It is easy to check that they form a symplectic basis for this level. The only remaining task to handle is  to complete Gram--Schmidt symplectic orthogonalization (sGS, in what follows) for the whole Jordan chain. The change of basis matrix provided by the components of the symplectic basis is a symplectic matrix, by construction, and provides us with the canonical form of the corresponding block.

If the eigenvalue is real, the generalised eigenvectors are real, and one carries out sGS similarly for each sector. We set aside the case of imaginary eigenvalues and that of eigenvalue zero, since they are relevant to the positive semidefinite case that we investigate in the next subsection.

\subsection{Positive semidefinite Hamiltonians}
\label{sec:posit-semid-hamilt}

Let us now address the case of interest to us here, that of positive semidefinite Hamiltonians. In this situation there are only three possible types of canonical symplectic blocks, namely i)  simple harmonic oscillators, with canonical form
\begin{equation}
	\label{eq:hocanonical}
	\mathsf{H}_{ho}=
	\begin{pmatrix}
		\omega&0\\0&\omega
	\end{pmatrix}\,,
\end{equation}
ii) free particles, with canonical form
\begin{equation}
	\label{eq:freecanonical}
	\mathsf{H}_f=
	\begin{pmatrix}
		0&0\\0&1
	\end{pmatrix}\,,
\end{equation}
and iii) nondynamical variables with identically zero blocks. 
The Hamiltonian will thus be symplectic diagonalizable. 

The issue for an effective algorithm applicable to the cases of interest is the disentangling of the free particle sector. It arises in the difference between the kernel of $\mathsf{JH}$, $K_1=\mathrm{ker}\left[\mathsf{JH}\right]$ and that of  $K_2=\mathrm{ker}\left[\left(\mathsf{JH}\right)^2\right]$. For this case of positive semidefinite $\mathsf{H}$ the generalized eigenspace of eigenvalue 0, $V_0$, is precisely $K_2$. In other words, the maximum length of a Jordan chain for this eigenvalue is two.

If $K_2=K_1$ there are no free particles, and $\mathsf{JH}$ is diagonalizable. In this situation the speediest way of achieving our goal of symplectic diagonalization of $\mathsf{H}$ in the relevant sector is to compute $\mathrm{ker}\left[\mathsf{H}\right]=K_1$, and thus the orthogonal decomposition $\mathbb{R}^{2n}=K_1\oplus K_1^\perp$. The restriction of $\mathsf{H}$ to $K_1^\perp$, denoted here as $\mathsf{H}_+$, is positive definite, and all the existing methods apply (see for instance section 8.3 of \cite{de2006symplectic}).

Let us examine the general case, which will provide us with one such method for the simpler situation as well. Define the linear subspace
\begin{equation}
	\label{eq:Edef}
	E = \mathsf{JH}\left[K_2\right]\subseteq K_1\,,
\end{equation}
and identify formally
\begin{equation}
	\label{eq:Fdef}
	\tilde{F} = \left(\mathsf{JH}\right)^{-1}\left[E\right]\,.
\end{equation}
To be definite, this means the following: obtain a basis of $E$, $\left\{\tilde{\mathbf{e}}_I\right\}_{I=1}^{n_f}$, and assign to each element of the basis its Jordan chain, i.e. a vector $\tilde{\mathbf{f}}_I$ such that $\mathsf{JH}\tilde{\mathbf{f}}_I= \tilde{\mathbf{e}}_I$. $\tilde{F}$ is the linear span of these vectors,
\begin{equation}
	\label{eq:Fexplicit}
	\tilde{F} = \mathrm{span}\left\{\tilde{\mathbf{f}}_I\right\}_{I=1}^{n_f}\,.
\end{equation}
Complete $\left\{\tilde{\mathbf{e}}_I\right\}_{I=1}^{n_f}$ to a basis of $K_1$ with a set of vectors $\left\{\tilde{\mathbf{w}}_i\right\}_{i=1}^{2n_{nd}}$, and
denote the span of these as  $\tilde{W}$
\begin{equation}
	\label{eq:Wdef}
	\tilde{W}=\mathrm{span}\left\{\tilde{\mathbf{w}}_i\right\}_{i=1}^{2n_{nd}}\,.
\end{equation}
It is important to notice that $\left\langle\tilde{\mathbf{w}}_i,\mathsf{J}\tilde{\mathbf{e}}_I\right\rangle=0$ ($\tilde{W}$ belongs to the symplectic complement of $E$), and, similarly,  $\left\langle\tilde{\mathbf{e}}_I,\mathsf{J}\tilde{\mathbf{e}}_J\right\rangle=0$ ($E$ is isotropic), because
\begin{align}
	\label{eq:wsymporthe}
	\left\langle\tilde{\mathbf{w}}_i,\mathsf{J}\tilde{\mathbf{e}}_I\right\rangle&= \left\langle\tilde{\mathbf{w}}_i,\mathsf{J}\left[\mathsf{JH}\tilde{\mathbf{f}}_I\right]\right\rangle\nonumber\\
	&= -\left\langle\tilde{\mathbf{w}}_i,\mathsf{H}\tilde{\mathbf{f}}_I\right\rangle\nonumber\\
	&= -\left\langle\mathsf{H}\tilde{\mathbf{w}}_i,\tilde{\mathbf{f}}_I\right\rangle=0\,.
\end{align}
We now carry out sGS for the free particle sector, making use of the fact that $E$ is isotropic, defining recursively for $I=1$ to $n_f$
\begin{align}
	\label{eq:sGSEF}
	\mathbf{e}_I &= a_I\left[\tilde{\mathbf{e}}_I-\sum_{J<I}\left\langle\tilde{\mathbf{e}}_I,\mathsf{J}\mathbf{f}_J\right\rangle \mathbf{e}_J\right]\,,\\
	\mathbf{f}_I &= a_I\left[\tilde{\mathbf{f}}_I-\sum_{J<I}\left\langle\tilde{\mathbf{f}}_I,\mathsf{J}\mathbf{f}_J\right\rangle \mathbf{e}_J+\sum_{J<I}\left\langle\tilde{\mathbf{f}}_I,\mathsf{J}\mathbf{e}_J\right\rangle \mathbf{f}_J\right]\,.\nonumber
\end{align}
The normalization factors are determined to be
\begin{equation}
	\label{eq:aIsquared}
	a_I^{-2}= \left\langle \tilde{\mathbf{f}}_I,\mathsf{H}\tilde{\mathbf{f}}_I\right\rangle - \sum_{J<I}\left\langle\tilde{\mathbf{e}}_I,\mathsf{J}\mathbf{f}_J\right\rangle\left\langle\mathbf{e}_J,\mathsf{J}\tilde{\mathbf{f}}_I\right\rangle\,.
\end{equation}
By induction one can prove that for all $I$ from 1 to $n_f$ $\mathsf{JH}\mathbf{f}_I=\mathbf{e}_I$, so the canonical Jordan form is preserved in the new basis for this sector.
Observe that $E$ is also the span of $\left\{\mathbf{e}_I\right\}_{I=1}^{n_f}$, and with the sGS we have constructed a linear space $F=\mathrm{span}\left\{\mathbf{f}_I\right\}_{I=1}^{n_f}$ that satisfies both the condition $\mathsf{JH}F=E$ and that of forming the symplectic subspace $W_F=E\oplus F$.
Now we want to obtain the symplectic complement of $W_F$ in $K_2$. To that purpose, we have to substract from $\tilde{W}$ the symplectic projection of $W_F$ on it, defining
\begin{equation}
	\label{eq:newW}
	\mathbf{w}_i = \tilde{\mathbf{w}}_i-\sum_{J=1}^{n_f}\left\langle\tilde{\mathbf{w}}_i,\mathsf{J}\mathbf{f}_J\right\rangle \mathbf{e}_J
\end{equation}
and their linear span $W=\mathrm{span}\left\{\mathbf{w}_i\right\}_{i=1}^{2n_{nd}}$. This is a symplectic subspace, and it is symplectic orthogonal to $W_f$ and to the harmonic oscillator sector.

The only complication in the harmonic oscillator sector might come from the presence of degenerate frequencies, i.e. from the presence of degenerate $i\omega_a$ eigenvalues of $\mathsf{JH}$. Let us denote with $\mathbf{v}^{(\alpha)}_a$ an eigenvector of $\mathsf{JH}$ for eigenvalue $i\omega_a$, where the superscript $\alpha$ is a degeneracy index. Then $\bar{\mathbf{v}}^{(\alpha)}_a$ is an eigenvector of $\mathsf{JH}$ for eigenvalue $-i\omega_a$. Furthermore, let $V_a$ be the (proper) eigenspace of $\mathsf{JH}$ for eigenvalue $i\omega_a$, and $\bar{V}_a$ its complex conjugate. Then $V_a\oplus \bar{V}_a$ is a symplectic subspace. Alternatively, we can write this as $\mathrm{Re}\left(V_a\right)\otimes\mathrm{Im}\left(V_a\right)$ as a real symplectic subspace. Thus the only remaining task is to construct a symplectic basis of this subspace, again by sGS. Namely, define
\begin{align}
	\label{eq:hopresymplectic}
	\tilde{\mathbf{e}}_a^{(\alpha)}& =\mathrm{Re}\left(\mathbf{v}_a^{(\alpha)}\right)\,,\nonumber\\
	\tilde{\mathbf{f}}_a^{(\alpha)}& =\mathrm{Im}\left(\mathbf{v}_a^{(\alpha)}\right)\,.
\end{align}
We perform this assignment because  $\left\langle\mathrm{Re}\left(\mathbf{v}_a^{(\alpha)}\right),\mathsf{J}\,\mathrm{Im}\left(\mathbf{v}_a^{(\alpha)}\right)\right\rangle$ is positive. This statement might seem surprising, since one can change the phase of $\mathbf{v}_a^{(\alpha)}$ {\it ad libitum}. However, as $\mathsf{JH}\mathbf{v}_a^{(\alpha)}= i\omega_a \mathbf{v}_a^{(\alpha)}$, it follows that
\begin{align}
	\label{eq:decompHOeigeneq}
	\mathsf{JH}\,\mathrm{Re}\left(\mathbf{v}_a^{(\alpha)}\right)&= -\omega_a\,\mathrm{Im}\left(\mathbf{v}_a^{(\alpha)}\right)\,,\nonumber\\
	\mathsf{JH}\,\mathrm{Im}\left(\mathbf{v}_a^{(\alpha)}\right)&= \omega_a\,\mathrm{Re}\left(\mathbf{v}_a^{(\alpha)}\right)\,,
\end{align}
whence
\begin{align}
	\label{eq:positivityproof}
	\omega_a\left\langle\tilde{\mathbf{e}}_a^{(\alpha)},\mathsf{J}\,\tilde{\mathbf{f}}_a^{(\alpha)}\right\rangle&=-\left\langle \mathrm{Re}\left(\mathbf{v}_a^{(\alpha)}\right),\tilde{\mathbf{e}}_a^{(\alpha)}\mathsf{J}\,\mathsf{JH}\left[\tilde{\mathbf{e}}_a^{(\alpha)}\right]\right\rangle\nonumber\\
	&= \left\langle\tilde{\mathbf{e}}_a^{(\alpha)},\mathsf{H}\, \tilde{\mathbf{e}}_a^{(\alpha)}\right\rangle >0\,.
\end{align}

Now, from this assignment, and as usual, define recursively
\begin{align}
	\label{eq:epsc}
	\mathbf{e}_a^{(a)}&= A_a^{(\alpha)}\left[\tilde{\mathbf{e}}_a^{(a)}-\sum_{\beta<\alpha}\left\langle\tilde{\mathbf{e}}_a^{(\alpha)},\mathsf{J}\mathbf{f}_a^{(\beta)}\right\rangle \mathbf{e}_a^{(\beta)}\right.\nonumber\\
	&\quad\left.+ \sum_{\beta<\alpha}\left\langle\tilde{\mathbf{e}}_a^{(\alpha)},\mathsf{J}\mathbf{e}_a^{(\beta)}\right\rangle \mathbf{f}_a^{(\beta)}\right]\,,
\end{align}
and similarly for $\mathbf{f}_a^{(\alpha)}$,
with the normalization selected to ensure that we obtain a symplectic basis. We choose the same factor $A_a^{(\alpha)}$ for both $\mathbf{e}_a^{(\alpha)}$ and $\mathbf{f}_a^{(\alpha)}$, to ensure that their combination $\mathbf{e}_a^{(\alpha)}+ i \mathbf{f}_a^{(\alpha)}$ is also an eigenvector of $\mathsf{JH}$ and therefore that the relations \eqref{eq:decompHOeigeneq} hold as
\begin{align}
	\label{eq:decompHOfinal}
	\mathsf{JH} \,\mathbf{e}_a^{(\beta)} &= -\omega_a \mathbf{f}_a^{(\beta)}\,,\nonumber\\
	\mathsf{JH} \,\mathbf{f}_a^{(\beta)} &= \omega_a \mathbf{e}_a^{(\beta)}\,.
\end{align}
We now prove this statement inductively. To do so, define
\begin{equation}
	\label{eq:uef}
	\mathbf{u}_a^{(\alpha)}=  \mathbf{e}_a^{(\alpha)}+ i \mathbf{f}_a^{(\alpha)}\,.
\end{equation}
If the inductive hypothesis  Eq. \eqref{eq:decompHOfinal} holds for $\beta<\alpha$, then for those $\beta<\alpha$
\begin{equation}
	\label{eq:jhubeta}
	\mathsf{JH}\,\mathbf{u}_a^{(\beta)}=i\omega_a \mathbf{u}_a^{(\beta)}\,,
\end{equation}
and therefore
\begin{equation}
	\label{eq:jhbarubeta}
	\mathsf{JH}\,\bar{\mathbf{u}}_a^{(\beta)}=-i\omega_a \bar{\mathbf{u}}_a^{(\beta)}\,.
\end{equation}
Consequently,
\begin{equation}
	\label{eq:jvbaru}
	\left\langle \mathbf{v}_a^{(\alpha)},\mathsf{J}\,\bar{\mathbf{u}}_a^{(\beta)}\right\rangle=0\,,
\end{equation}
as follows from Eq. \eqref{eq:symporth}, from which one can conclude that
\begin{align}
	\label{eq:internalrel}
	\left\langle \tilde{\mathbf{e}}_a^{(\alpha)},\mathsf{J}\,{\mathbf{e}}_a^{(\beta)}\right\rangle -
	\left\langle \tilde{\mathbf{f}}_a^{(\alpha)},\mathsf{J}\,{\mathbf{f}}_a^{(\beta)}\right\rangle&=0\,,\nonumber\\
	\left\langle \tilde{\mathbf{e}}_a^{(\alpha)},\mathsf{J}\,{\mathbf{f}}_a^{(\beta)}\right\rangle +
	\left\langle \tilde{\mathbf{f}}_a^{(\alpha)},\mathsf{J}\,{\mathbf{e}}_a^{(\beta)}\right\rangle&=0      
\end{align}
for $\beta<\alpha$ if the inductive hypothesis holds. Clearly Eq. \eqref{eq:decompHOfinal} holds for $\beta=1$, the first inductive step. It is now easy to check that if it holds for $\beta<\alpha$, then Eq. \eqref{eq:epsc} and its $\mathbf{f}$ companion allow us to conclude that \eqref{eq:decompHOfinal} also holds for $\beta=\alpha$.

Now we complete our task. The matrix constructed as
\begin{equation}
	\label{eq:sfinal}
	\mathsf{S} =
	\begin{pmatrix}
		\left\{\mathbf{w}_i\right\}_{i=1}^{2 n_{nd}} & \left\{\mathbf{e}_I\right\}_{I=1}^{n_f} & \left\{\mathbf{f}_I\right\}_{I=1}^{n_f} & \left\{\mathbf{e}_a^{(\alpha)}\right\} &\left\{\mathbf{f}_a^{(\alpha)}\right\}
	\end{pmatrix}
\end{equation}
is not necessarily symplectic, because the first set, $ \left\{\mathbf{w}_i\right\}_{i=1}^{2 n_{nd}}$, need not be a symplectic basis. If insisted upon, sGS can also be applied to the nondynamical sector, and after the procedure $\mathsf{S}$ will indeed be symplectic (up to reordering, as we shall see). But even without going through that step, it is indeed the case that
\begin{equation}
	\label{eq:sths}
	\mathsf{S}^T \mathsf{H} \mathsf{S} = \mathsf{H}_D = \mathrm{diag}
	\begin{pmatrix}
		0&0&\mone&\msOmega&\msOmega
	\end{pmatrix}\,,
\end{equation}
where $\msOmega$ is the diagonal matrix constructed with $\omega_a$ with its possible degeneracies.

Let us prove this statement, making use of all the previous analysis. To do so, let us use a more compact notation introducing the rectangular matrices $\mathsf{W}$, $\mathsf{E}$, $\mathsf{F}$, $\mathcal{E}$ and $\mathcal{F}$ whose columns are, respectively, the vectors $\mathbf{w}_i$, $\mathbf{e}_I$, $\mathbf{f}_I$, $\mathbf{e}_a^{(\alpha)}$ and $\mathbf{f}_a^{(\alpha)}$. Bearing in mind that $\mathsf{J}$ is invertible and considering the action of $\mathsf{JH}$ on each of this rectangular matrices, we have
\begin{align}
	\label{eq:listofactions}
	\mathsf{H}\,\mathsf{W}&=0\,
	\mathsf{H}\,\mathsf{E}=0\,,
	\mathsf{H}\,\mathsf{F}=-\mathsf{J}\,\mathsf{E}\,,\\
	\mathsf{H}\,\mathcal{E}&=\mathsf{J}\,\mathcal{F}\mathsf{\Omega}\,,
	\mathsf{H}\,\,\mathcal{\mathcal{F}}=-\mathsf{J}\,\mathcal{E}\mathsf{\Omega}\,.\nonumber
\end{align}
The first two are simply the statement that the $\mathbf{w}_i$ and $\mathbf{e}_I$ vectors belong to the kernel of $\mathsf{H}$. The third line  that $\mathsf{F}$ and $\mathsf{E}$ are related as elements of a Jordan chain. The last two lines are actually a restatement of Eq. \eqref{eq:decompHOfinal}.

We now see that
\begin{align}
	\label{eq:hons}
	\mathsf{H}\,\mathsf{S} &= \mathsf{H}
	\begin{pmatrix}
		\mathsf{W}&\mathsf{E}&\mathsf{F}&\mathcal{E}&\mathcal{F}
	\end{pmatrix}\nonumber\\
	&=
	\begin{pmatrix}
		0&0&-\mathsf{J}\,\mathsf{E}& \mathsf{J}\mathcal{F}\mathsf{\Omega}&-\mathsf{J}\,\mathcal{E}\mathsf{\Omega}
	\end{pmatrix}\,.
\end{align}
From here we compute Eq. \eqref{eq:sths} explicitly:
\begin{align}
	\label{eq:sthsexplicit}
	\mathsf{S}^T\mathsf{HS}&=
	\begin{pmatrix}
		\mathsf{W}^T\\ \mathsf{E}^T\\ \mathsf{F}^T\\ \mathcal{E}^T\\ \mathcal{F}^T
	\end{pmatrix}\begin{pmatrix}
		0&0&-\mathsf{J}\,\mathsf{E}& \mathsf{J}\mathcal{F}\mathsf{\Omega}&-\mathsf{J}\,\mathcal{E}\mathsf{\Omega}
	\end{pmatrix}\nonumber\\
	&=
	\begin{pmatrix}
		0&0&-\mathsf{W}^T\mathsf{J}\,\mathsf{E}& \mathsf{W}^T\mathsf{J}\mathcal{F}\mathsf{\Omega}&-\mathsf{W}^T\mathsf{J}\,\mathcal{E}\mathsf{\Omega}\\
		0&0&-\mathsf{E}^T\mathsf{J}\,\mathsf{E}& \mathsf{E}^T\mathsf{J}\mathcal{F}\mathsf{\Omega}&-\mathsf{E}^T\mathsf{J}\,\mathcal{E}\mathsf{\Omega}\\
		0&0&-\mathsf{F}^T\mathsf{J}\,\mathsf{E}& \mathsf{F}^T\mathsf{J}\mathcal{F}\mathsf{\Omega}&-\mathsf{F}^T\mathsf{J}\,\mathcal{E}\mathsf{\Omega}\\
		0&0&-\mathcal{E}^T\mathsf{J}\,\mathsf{E}& \mathcal{E}^T\mathsf{J}\mathcal{F}\mathsf{\Omega}&-\mathcal{E}^T\mathsf{J}\,\mathcal{E}\mathsf{\Omega}\\
		0&0&-\mathcal{F}^T\mathsf{J}\,\mathsf{E}& \mathcal{F}^T\mathsf{J}\mathcal{F}\mathsf{\Omega}&-\mathcal{F}^T\mathsf{J}\,\mathcal{E}\mathsf{\Omega}
	\end{pmatrix}\\
	&=
	\begin{pmatrix}
		0&0&0&0&0\\
		0&0&0&0&0\\
		0&0&\mone&0&0\\
		0&0&0&\mathsf{\Omega}&0\\
		0&0&0&0&\mathsf{\Omega}\\
	\end{pmatrix}=\mathsf{H}_D\,.
\end{align}
In the first row the third zero follows from Eq. \eqref{eq:wsymporthe}. The fourth and the fifth from that row, as well as from the second and third one, are a consequence of Eq. \eqref{eq:symporth}. Similarly for the third zero of the fourth and fifth rows. The third zero of the second row is from the analogue of Eq. \eqref{eq:wsymporthe} in the $E$ subspace. The identity in the third element of the third row follows from the fact that $\left\{\mathbf{e}_I, \mathbf{f}_I\right\}_{I=1}^{n_f}$ is a symplectic basis. And analogously for the harmonic oscillator block.

Regarding whether this matrix $\mathsf{S}$ is symplectic or not, we can compute $\mathsf{S}^T\mathsf{JS}$ explicitly, using the same arguments, and it results in
\begin{equation}
	\label{eq:stjs}
	\mathsf{S}^T\mathsf{JS}=
	\begin{pmatrix}
		\mathsf{W}^T\mathsf{JW}&0&0&0&0\\
		0&0&\mone&0&0\\
		0&-\mone&0&0&0\\
		0&0&0&0&\mone\\
		0&0&0&-\mone&0
	\end{pmatrix}\,.
\end{equation}
We are guaranteed that $\mathsf{W}^T\mathsf{JF}=0$ because we have reorganized $\tilde{W}$ into $W$ in Eq. \eqref{eq:newW}. The other entries are due to sGS in the different sectors and Eq. \eqref{eq:wsymporthe}. We are not overly concerned with the presence of the antisymmetric entry  $\mathsf{W}^T\mathsf{JW}$, because it pertains to the nondynamical sector which will be eliminated from any posterior analysis. To make this statement abundantly clear, consider the classical equations of motion for the initial variables, Eq. \eqref{eq:eoms_phase_space_H}, that we restate here:
\begin{align*}
	\dot{\bx}=\msJ\msH\bx\,.
\end{align*}
Let us define now $\by$ by $\bx=\msS\by$, with $\msS$ defined in Eq. \eqref{eq:sfinal}, i.e.
\begin{equation}
	\label{eq:sblock}
	\mathsf{S}=\begin{pmatrix}
		\mathsf{W}&\mathsf{E}&\mathsf{F}&\mathcal{E}&\mathcal{F}
	\end{pmatrix}\,.
\end{equation}
From Eq. \eqref{eq:stjs} one can determine its inverse,
\begin{equation}
	\label{eq:sinverse}
	\mathsf{S}^{-1}=
	\begin{pmatrix}
		\left(\mathsf{W}^T\mathsf{JW}\right)^{-1}&0&0&0&0\\
		0&0&-\mone&0&0\\
		0&\mone&0&0&0\\
		0&0&0&0&-\mone\\
		0&0&0&\mone&0
	\end{pmatrix}\mathsf{S}^T\mathsf{J}\,. 
\end{equation}
Thus
\begin{align}
	\label{eq:yeoms}
	\dot{\by}&= \mathsf{S}^{-1}\mathsf{JH}\,\mathsf{S}\by \nonumber\\
	&=
	\begin{pmatrix}
		0&0&0&0&0\\
		0&0&\mone&0&0\\
		0&0&0&0&0\\
		0&0&0&0&\mathsf{\Omega}\\
		0&0&0&-\mathsf{\Omega}&0
	\end{pmatrix}\by\,,
\end{align}
as promised throughout the analysis. The possibly nontrivial $\mathsf{W}^T\mathsf{JW}$ plays no role, as it pertains to the nondynamical sector.

\begin{widetext}
\section{Landau quantization}
\label{App:Landau_quantization}
Working on the symmetric gauge, we can write the Hamiltonian for the two linearly-coupled charged particles in a magnetic field ($z$ direction) as $H=\frac{1}{2}\bX_z^T\msH_z\bX_z+\frac{1}{2}\bX_{xy}^T\msH_{xy}\bX_{xy}$ where 
\begin{align}
	\msH_z&=\begin{pmatrix}
		k&-k&0&0\\-k&k&0&0\\0&0&\frac{1}{m}&0\\0&0&0&\frac{1}{m}
	\end{pmatrix},\label{eq:Hz_Landau}\\
	\msH_{xy}&=\begin{pmatrix}
		\tilde{k}&-k&0&0&0&0&\frac{-g}{m}&0\\
		-k&\tilde{k}&0&0&0&0&0&\frac{-g}{m}\\
		0&0&\tilde{k}&-k&\frac{g}{m}&0&0&0\\
		0&0&-k&\tilde{k}&0&\frac{g}{m}&0&0\\
		0&0&\frac{g}{m}&0&\frac{1}{m}&0&0&0\\
		0&0&0&\frac{g}{m}&0&\frac{1}{m}&0&0\\
		\frac{-g}{m}&0&0&0&0&0&\frac{1}{m}&0\\
		0&\frac{-g}{m}&0&0&0&0&0&\frac{1}{m}
	\end{pmatrix}.\label{eq:Hxy_Landau}
\end{align}
with $g=eB/2$, $\tilde{k}=k+\frac{g^2}{m}$, $\bX_z=(z_1,z_2,p_{z,1},p_{z,2})^T$ and $\bX_{xy}=(x_1,x_2,y_1,y_2,p_{x,1},p_{x,2},p_{y,1},p_{y,2})^T$. The harmonic oscillator in the $z$ direction has frequency $\Omega_z=\sqrt{2k/m}$.  Now, let us show how to perform a blindfold diagonalization of the dynamics in the $x$--$y$. By first making a rescaling of the variables to simplify the parameters, the configuration space variables in the plane are modified as $q_i\rightarrow\frac{q_i}{\sqrt{m\Omega_c}}$ and the momenta accordingly $p_i\rightarrow \sqrt{m\Omega_c}p_i$, where $\Omega=g/2m=\Omega_c/4$, with $\Omega_c=eB/m$ the cyclotron frequency. This Hamiltonian can thus be written as 

\begin{align}
	\msH_{xy}&=\Omega_c
	\begin{pmatrix}
		\chi ^2+\frac{1}{4} & -\chi ^2 & 0 & 0 & 0 & 0 & -\frac{1}{2} & 0 \\
		-\chi ^2 & \chi ^2+\frac{1}{4} & 0 & 0 & 0 & 0 & 0 & -\frac{1}{2} \\
		0 & 0 & \chi ^2+\frac{1}{4} & -\chi ^2 & \frac{1}{2} & 0 & 0 & 0 \\
		0 & 0 & -\chi ^2 & \chi ^2+\frac{1}{4} & 0 & \frac{1}{2} & 0 & 0 \\
		0 & 0 & \frac{1}{2} & 0 & 1 & 0 & 0 & 0 \\
		0 & 0 & 0 & \frac{1}{2} & 0 & 1 & 0 & 0 \\
		-\frac{1}{2} & 0 & 0 & 0 & 0 & 0 & 1 & 0 \\
		0 & -\frac{1}{2} & 0 & 0 & 0 & 0 & 0 & 1
	\end{pmatrix},\label{eq:H_xy_simp_Landau}
\end{align}
where we have defined the unitless parameter $\chi=\Omega_z/\Omega\sqrt{2}$. The normal frequencies of this sector can be found by  diagonalizing $\msJ_{xy}\msH_{xy}$, and correspond to the cyclotron motion ($\Omega$), and one symmetric and one antisymmetric modes  $\left(\Omega\sqrt{\sqrt{1+8 \chi ^2}\pm(1+4 \chi^2)/2}\right)$. Following the algorithm, we can obtain systematically symplectic transformations that diagonalize Hamiltonian (\ref{eq:H_xy_simp_Landau}). For example, at the degeneracy point $\chi=1$, 
\begin{align}
	\msS|_{\chi=1}=\begin{pmatrix}
		\msW & \mcl{E} &\mcl{F}
	\end{pmatrix}=\left(
	\begin{array}{cccccccc}
		\frac{1}{\sqrt{2}} & 0 & -\frac{1}{\sqrt{2}} & -\frac{1}{\sqrt{6}} & \frac{1}{\sqrt{6}} & 0 & 0 & 0 \\
		\frac{1}{\sqrt{2}} & 0 & -\frac{1}{\sqrt{2}} & \frac{1}{\sqrt{6}} & -\frac{1}{\sqrt{6}} & 0 & 0 & 0 \\
		0 & -\frac{1}{\sqrt{2}} & 0 & 0 & 0 & -\frac{1}{\sqrt{2}} & \frac{1}{\sqrt{6}} & \frac{1}{\sqrt{6}} \\
		0 & -\frac{1}{\sqrt{2}} & 0 & 0 & 0 & -\frac{1}{\sqrt{2}} & -\frac{1}{\sqrt{6}} & -\frac{1}{\sqrt{6}} \\
		0 & \frac{1}{2 \sqrt{2}} & 0 & 0 & 0 & -\frac{1}{2 \sqrt{2}} & - \sqrt{\frac{3}{8}} &  \sqrt{\frac{3}{8}} \\
		0 & \frac{1}{2 \sqrt{2}} & 0 & 0 & 0 & -\frac{1}{2 \sqrt{2}} &  \sqrt{\frac{3}{8}} & - \sqrt{\frac{3}{8}} \\
		\frac{1}{2 \sqrt{2}} & 0 & \frac{1}{2 \sqrt{2}} & - \sqrt{\frac{3}{8}} & - \sqrt{\frac{3}{8}} & 0 & 0 & 0 \\
		\frac{1}{2 \sqrt{2}} & 0 & \frac{1}{2 \sqrt{2}} &  \sqrt{\frac{3}{8}} &  \sqrt{\frac{3}{8}} & 0 & 0 & 0 \\
	\end{array}
	\right),
\end{align}

such that the diagonalized Hamiltonian becomes 
\begin{align}
\msH_D|_{\chi=1}=\Omega\left(
	\begin{array}{cccccccc}
		0 & 0 & 0 & 0 & 0 & 0 & 0 & 0 \\
		0 & 0 & 0 & 0 & 0 & 0 & 0 & 0 \\
		0 & 0 & 1 & 0 & 0 & 0 & 0 & 0 \\
		0 & 0 & 0 & 1 & 0 & 0 & 0 & 0 \\
		0 & 0 & 0 & 0 & 2 & 0 & 0 & 0 \\
		0 & 0 & 0 & 0 & 0 & 1 & 0 & 0 \\
		0 & 0 & 0 & 0 & 0 & 0 & 1 & 0 \\
		0 & 0 & 0 & 0 & 0 & 0 & 0 & 2 \\
	\end{array}
	\right).
	\end{align}
      \end{widetext}

\section{Conserved linear and quadratic quantities}
\label{sec:cons-line-quadr}
In any study of the Landau problem one will immediately encounter a description of the linear and quadratic conserved quantities. For the one particle Landau problem, the latter is the angular momentum, which reduces to the energy minus a constant associated with the nondynamical sector when on-shell. The characterization and identification of all linear and quadratic conserved quantities for a linear system is, as that example shows, useful and relevant, most importantly with a view towards the nonlinear case. In this section we observe that the application of  Williamson's theorem to  positive semidefinite Hamiltonians allows us to give a complete set of independent linear and quadratic conserved quantities.

The idea is the following: as there is a canonical symplectically diagonalized form of the Hamiltonian matrix, characterize the invariants in that canonical form, and undo the symplectic transformation to write the full set in the initial variables. We shall carry out the analysis using the classical equations of motion, because for  operators linear and quadratic in the initial variables the classical equations match Heisenberg's equations of motion.

The classical equations of motion for functions on phase space without explicit time dependence are
\begin{equation}
	\label{eq:poisson}
	\dot{f} = \left\{f,H\right\}\,,
\end{equation}
which for the case under consideration become
\begin{equation}
	\label{eq:eomf}
	\dot{f}= \mathsf{J}_{\alpha\beta}\partial_\alpha f\,\mathsf{H}_{\beta\gamma} x^\gamma\,.
\end{equation}
Alternatively,
\begin{equation}
	\label{eq:emofvecs}
	\dot{f}= -\bx^T\mathsf{H}\mathsf{J}\nabla f\,.
\end{equation}
If $f$ is a linear function on phase space, $f(\bx)=\bx^T\mathbf{f}+d$, with $\mathbf{f}$ a column vector, and thus the condition of being a conserved quantity is that for all points of phase space $\bx$ one has
\begin{equation}
	\label{eq:linearcondition}
	\bx^T\mathsf{H}\mathsf{J}\mathbf{f}=0\,.
\end{equation}
It follows that $\mathsf{J}\mathbf{f}$ must belong to the kernel of $\mathsf{H}$. Thus we identify the set of linear conserved quantities with $\mathsf{J}\mathrm{ker}\left[\mathsf{H}\right]$. If the reader is wondering about the need for the symplectic matrix, please observe, first, that it is invertible, so we indeed have that a linear space of linear conserved quantities that is isomorphic to the kernel of $\mathsf{H}$ and, second, that we have seen that the $\mathrm{ker}\left[\mathsf{H}\right]$ consists of the nondynamical sector of phase space and the position like part of the phase space of free particles, since it is the $E$ space. The conserved quantity for a free particle is its momentum, i.e. $F$ space in the notation above, and this the result from applying $\mathsf{J}$ to $E$, essentially.

Coming now to quadratic conserved quantities, we write the symmetric matrix associated with one such $f$ as $\mathsf{B}=\mathsf{B}^T$ (to avoid confusion with a previous use of $\mathsf{F}$), as
\begin{equation}
	\label{eq:quaddef}
	f(\bx)=\frac{1}{2}\bx^T\mathsf{B}\bx\,.
\end{equation}
Now Eq. \eqref{eq:emofvecs} becomes
\begin{equation}
	\label{eq:feqquad}
	\dot{f}= -\bx^T\mathsf{H}\mathsf{J}\mathsf{B}\bx\,.
\end{equation}
For $f$ to be a conserved quantity we need the right hand side to vanish for all points on phase space. Therefore the symmetric part of the matrix must be zero, that is
\begin{equation}
	\label{eq:quadconscond}
	\mathsf{HJB}-\mathsf{BJH}=0\,,
\end{equation}
which is the condition we must now study, given a Hamiltonian matrix $\mathsf{H}$. In fact, premultiplying with the symplectic matrix $\mathsf{J}$ one obtains the satisfactory physical interpretation that the evolution in time controlled by $\mathsf{H}$ commutes with that generated by $\msB$. Quantum mechanically, the associated Hamiltonians must conmute, and in terms of the matrices we have
\begin{equation}
	\label{eq:quadcondcomm}
	\left[\mathsf{JH},\mathsf{JB}\right]=0\,.
\end{equation}
Thus the problem is reduced to the study of the commutant of the normal form of $\mathsf{JH}$. If $\mathsf{H}$ is positive, the normal form is $\mathrm{diag}\left(i\Omega,-i\Omega\right)$, and is obtained from $\mathsf{JH}$ by conjugation. If there are no degeneracies, the commutant is given by diagonal matrices. A further condition, however, must be met, namely that $\mathsf{B}$ be real and  symmetric. Given $\mathbf{v}_a$ eigenvector of $\mathsf{JH}$ with eigenvalue $-i\omega_a$, it must be the case that it be an eigenvector of $\mathsf{JB}$, with eigenvalue $\beta_a$. It follows that $\mathbf{v}_a^*$ must also be an eigenvector of $\mathsf{JB}$, now with eigenvalue $\beta^*_a$. In terms of the real and imaginary parts of $\mathbf{v}_a$, $\mathbf{v}_a= \mathbf{e}_a+i \mathbf{f}_a$, and denoting the real and imaginary parts of $\beta_a$ as $\beta_a=\mu_a+i\nu_a$, we have that in the real symplectic  basis $\left\{\mathbf{e}_a,\mathbf{f}_a\right\}$ (or proportional to symplectic) $\mathsf{JB}$ has the matrix
\begin{equation}
	\label{eq:jba}
	\mathsf{JB}\Big|_{a}=
	\begin{pmatrix}
		\mu_a& \nu_a\\-\nu_a&\mu_a
	\end{pmatrix}\,,
\end{equation}
and thus, in the same basis
\begin{equation}
	\label{eq:ba}
	\msB_a =
	\begin{pmatrix}
		\nu_a&-\mu_a\\ \mu_a&\nu_a
	\end{pmatrix}\,.
\end{equation}
Since $\msB$ is symmetric, this means that (as expected) $\mu_a=0$.

Let us now examine the positive definite degenerate case, and use the notation of Eq. \eqref{eq:uef} in Section \ref{sec:posit-semid-hamilt}.  The matrix $\msB$ restricted to the $\omega_a$ eigenspace in the basis $\left\{e_a^{(\alpha)}\right\}\cup\left\{f_a^{(\alpha)}\right\}$ will be of the form 
\begin{equation}
	\label{eq:badeg}
	\msB_a =
	\begin{pmatrix}
		\msN_a&-\msM_a\\ \msM_a&\msN_a
	\end{pmatrix}
\end{equation}
with $\mathsf{M}_a$ and $\mathsf{N}_a$ matrices. As $\mathsf{B}$ is symmetric, this means that $\mathsf{M}_a$ must be antisymmetric and $\mathsf{N}_a$ symmetric.

For example, two harmonic oscillators with different frequencies have no quadratic invariants other than their separate energy functions and linear superpositions thereof. On the other hand the isotropic harmonic oscillator on the plane presents a basis of  four quadratic invariants: the two energy functions $\left(x_i^2+p_i^2\right)$, the canonical angular momentum $x_1p_2-x_2p_1$, and a final $x_1x_2+p_1p_2$ (or alternatively $\left(x_1+x_2\right)^2+\left(p_1+p_2\right)^2$).

Before examining the general positive semidefinite case, let us illustrate the difference with the  one particle Landau problem. After some symplectic transformations, the Hamiltonian matrix is proportional to
\begin{equation}
	\label{eq:hmatLandau1}
	\mathsf{H}=
	\begin{pmatrix}
		1&0&0&1\\0&1&-1&0\\0&-1&1&0\\ 1&0&0&1
	\end{pmatrix}\,,
\end{equation}
with one non dynamical degree of freedom and one harmonic oscillator. Thus the canonical form is proportional to
\begin{equation}
	\label{eq:hdlandau}
	\tilde{\mathsf{H}}_D=\mathrm{diag}
	\begin{pmatrix}
		0&1&0&1
	\end{pmatrix}
	\,.
\end{equation}
 On inserting this form in Eq. \eqref{eq:quadcondcomm} one obtains that in the canonical basis the quadratic invariants are the trivial ones involving the nondynamical sector (three independent ones, $\xi^2$, $p_\xi^2$, and $\xi p_\xi$, with $\xi,p_\xi$ canonical coordinates of the nondynamical sector), and the energy invariant of the oscillator, $x^2+p_x^2$.

 In this example one sees a feature of the general case: the harmonic oscillator sector will not couple with the nondynamical and free particle sectors in invariant quadratic forms. On the other hand, as can readily be seen from the trivial example of one free particle and one nondynamical, these do couple into invariants. These invariants, however, are in that sector not independent from the linear invariants, that generate all possible invariants. In summary, the only independent quadratic invariants are to be found in the harmonic oscillator sector.

 We can rephrase the presentation above in more abstract terms as follows: (i) $\mathsf{JH}$ belongs to the Lie algebra for the symplectic group; (2) By using symplectic transformations on $\mathsf{H}$, $\mathsf{JH}$ can be written as being upper triangular; (3) the quadratic conserved quantities are determined by the commutant of $\mathsf{JH}$ in the Lie algebra.

 \subsection{Gauge invariance and symplectic diagonalization}
\label{sec:gauge-invar-sympl}
The physical content of Hamiltonians such as  the minimal coupling example of Eq. \eqref{eq:MinimalCoupling} is independent of the gauge in which it is written. However, there is a limited set of gauges in which the Hamiltonian is a quadratic function on phase space. Namely, the symmetric gauge and those obtained from the symmetric gauge vector potential and adding the gradient of a quadratic function,
\begin{align}
  \label{eq:gauge}
  \bsb{A} = \bsb{A}_S+\nabla\xi
\end{align}
with
\begin{align}
  \label{eq:xi}
  \xi = \frac{1}{2}\bx^T \mathsf{\Xi} \bx\,.
\end{align}
In fact, this induces a symplectic transformation.

As the symplectic diagonalization approach is precisely a tool to identify the evolution of a system in terms of its symplectic invariants, changes of gauge such as these have no impact on the process and the result.
\section{Linear reciprocal networks: counting of degrees of freedom}
\label{app:LC_networks}

The Hamiltonian in Eq. \eqref{eq:H_LC_network} is evidently positive semidefinite. The corresponding Hamiltonian matrix is
\begin{align}
	\label{eq:hmatrixLC}
\mathsf{H}=\begin{pmatrix}
	0&0&0& 0\\
	0&\mathsf{D}^T\mathsf{L}^{-1}\mathsf{D}&-\mathsf{D}^T \mathsf{L}^{-1}&0\\
	0&-\mathsf{L}^{-1}\mathsf{D}& \msL^{-1}&0\\
	0&0&0&\msC^{-1}
\end{pmatrix}\,.
\end{align}
$\mathsf{C}$ and $\mathsf{L}$ are positive definite matrices defined on vectors with $N_C$ and $M_L$ components respectively. The adjacency matrix $\mathsf{D}$ has dimension $N_C\times M_L$ and is possibly not of maximum rank.
Denote with $\tilde{\be}$ a generic element of the kernel of $\mathsf{D}$ and with $\tilde{\be}_T$  a generic element of $\mathrm{ker}\left[\mathsf{D}^T\right]$. Similarly represent a generic element of $\mathbb{R}^{N_C}$ as $\bphi$, and a generic element of $\mathbb{R}^{M_L}$ as $\boldsymbol{q}$.

In order to assess the number of nondynamical, free particle and harmonic oscillator degrees of freedom we have to compute the dimensions of $K_1=\mathrm{ker}\left[\mathsf{JH}\right]$ and $K_2=\mathrm{ker}\left[\left(\mathsf{JH}\right)^2\right]$.  In this case it is easy to compute an explicit form for the generic elements of $K_1$ and $K_2$, namely
\begin{align}\label{eq:k1andk2linear}
	\begin{pmatrix}
		\bq\\ \bphi\\ \msD\bphi\\ 0
	\end{pmatrix}\in K_1,\qquad\text{and}\qquad
	\begin{pmatrix}
		\bq\\ 
		\bphi\\ 
		\msD\bphi+\msL\tilde{\be}_T\\ 
		\msC\tilde{\be}
	\end{pmatrix}\in K_2\,.
\end{align}

We read directly the dimensions and the number of free particles,
\begin{align}
	\label{eq:simplecounting}
	\mathrm{dim}\left(K_1\right)&= N_C+M_L\,,\nonumber\\
	\mathrm{dim}\left(K_2\right) &= N_C+M_L+n_f\,,\nonumber\\
	n_f&= \mathrm{dim}\left(\mathrm{ker}\left[\mathsf{D}\right]\right)+\mathrm{dim}\left(\mathrm{ker}\left[\mathsf{D}^T\right]\right)\,.
\end{align}
It is easy to build $E$ and $\tilde{F}$ spaces, in the notation of Appendix \ref{sec:sympl-diag-posit}, as being generated, respectively by vectors of the form
\begin{align}\label{eq:EandtildeFlinear}
  \begin{pmatrix}
    \tilde{\be}_T\\ \tilde{\be}\\0\\0
  \end{pmatrix}\in E\qquad \mathrm{and}\quad
  \begin{pmatrix}
    0\\0\\ \mathsf{L}\tilde{\be}_T\\ \mathsf{C}\tilde{\be}
  \end{pmatrix}\in \tilde{F}\,.
\end{align}
Since the dimension of $K_1$ is the number of free particles plus twice the number of nondynamical variables, we express the number of nondynamical variables for this case of linear reciprocal networks as
\begin{align}
	\label{eq:ndspecial}
	n_{nd}&= \frac{1}{2}\left(\mathrm{dim}\left(K_1\right)-n_f\right)\nonumber\\
	&= \frac{1}{2}\left[N_C+M_L-n_f\right]\,.
\end{align}
Passing now to the harmonic oscillator sector, notice that $n_{ho}+n_f+n_{nd}$ is the total number of configuration space variables, in our case $N_C+M_L$, whence we compute
\begin{align}
	\label{eq:nhospecial}
	n_{ho}&= N_C+M_L- n_f-n_{nd}\nonumber\\
	&= \frac{1}{2}\left[N_C+M_L-n_f\right]=n_{nd}\,,
\end{align}
so the number of harmonic oscillators is the same as that of nondynamical variables in this case of linear reciprocal networks, as stated in the main text.

\subsection{Example}
\label{sec:example}

In the text the example of Fig. \ref{fig:LCC_circuit} (b) has been put forward. In this situation, with $\mathsf{D}=
\begin{pmatrix}
  1&1
\end{pmatrix}$, we see that
$\mathrm{ker}\left[\mathsf{D}^T\right]=\emptyset$, while  $\mathrm{ker}\left[\mathsf{D}\right]$ is one-dimensional, so $n_f=1$. We have $N_C=2$, $M_L=1$, whence $n_{nd}=n_{ho}=1$.
Applying Eqs. \eqref{eq:sGSEF} and \eqref{eq:aIsquared}, we obtain for the free particle sector
\begin{align}
  \label{eq:effreeexample}
  \mathbf{e}_f =\frac{1}{\sqrt{C_1+C_2}}
  \begin{pmatrix}
    0\\1\\-1\\0\\0\\0
  \end{pmatrix}\quad\mathrm{and}\quad \mathbf{f}_f=\frac{1}{\sqrt{C_1+C_2}}
  \begin{pmatrix}
    0\\0\\0\\0\\C_1\\ - C_2
  \end{pmatrix}\,.
\end{align}
From inspection of $K_1$ in \eqref{eq:k1andk2linear} and comparison \eqref{eq:effreeexample} we construct $\tilde{W}$, as defined in Eq. \eqref{eq:Wdef}, and by symplectic orthogonalization, Eq. \eqref{eq:newW} we obtain
\begin{align}
  \label{eq:wexample}
  \mathbf{w}_1=
  \begin{pmatrix}
    1\\0\\0\\0\\0\\0
  \end{pmatrix}\,,\qquad \mathbf{w}_2=\frac{1}{C_1+C_2}
  \begin{pmatrix}
    0\\ C_2\\ C_1\\ C_1+C_2\\0\\0
  \end{pmatrix}\,.
\end{align}
The full symplectic basis is completed with the harmonic oscillator sector,
\begin{align}
  \label{eq:hoexample}
  \mathbf{e}= \frac{1}{\sqrt{\Omega L}}
  \begin{pmatrix}
    1\\0\\0\\1\\1
  \end{pmatrix}\,,\quad\mathbf{f}= \frac{-1}{\Omega^{3/4}L^{1/2}}
  \begin{pmatrix}
    0\\1/C_1\\1/C_2\\0\\0\\0
  \end{pmatrix},\,
\end{align}
where $\Omega=1/\sqrt{L(C_1+C_2)}$.
%
 
	\begin{widetext}
	\section{Example circuit: JJs coupled to a NR black-box admittance}
	\label{App:JJs_Cc_Ymat}
	Here we provide details for the quantization of the nonreciprocal two-port admittance capacitively coupled to Josephson junctions in Fig.~\ref{fig:2CQ_2_port_NR_Y}. After writing systematically the Lagrangian (\ref{eq:L_blackbox}) and Hamiltonian (\ref{eq:H_JJs_Cc_Ymat}), let us perform a triangular transformation (shift) of the internal charge degrees of freedom with the Josephson fluxes,
	\begin{align}
		\tilde{\bQ}&=\bQ-2\msZ^{-1}\msD_J\bPhi_J,\quad	 \tilde{\bP}=\bP,\\
		\tilde{\bPhi}_J&=\bPhi_J,\quad\tilde{\bPi}_J=\bPi_J-2\msD_J^T\msZ^{-1}\bP,
	\end{align}
	while not modiying the coupling capacitor variables $\tilde{\bPhi}_c=\bPhi_c$ and $\tilde{\bPi}_c=\bPi_c$. The Hamiltonian can be rewritten then
	\begin{align}
		H=&\,\frac{1}{2}(\tilde{\bP}-\tfrac{1}{2}\msZ\tilde{\bQ}+\msD_c\tilde{\bPhi}_c)^T\msL^{-1}(\tilde{\bP}-\tfrac{1}{2}\msZ\tilde{\bQ}+\msD_c\tilde{\bPhi}_c)\nonumber\\
		&+\frac{1}{2}(\tilde{\bPi}_J+2\msD_J^T\msZ^{-1}\tilde{\bP})^T\msC_J^{-1}(\tilde{\bPi}_J+2\msD_J^T\msZ^{-1}\tilde{\bP})+\frac{1}{2}\tilde{\bPi}_c\msC_c^{-1}\tilde{\bPi}_c+U(\tilde{\bPhi}_J)\nonumber\\
		=&\tilde{\bX}^T\msH\tilde{\bX}+U(\tilde{\bPhi}_J)\nonumber\\
		=&\frac{1}{2}\left[\tilde{\bY}^T\msH_{\text{lin}}\tilde{\bY}+\tilde{\bPi}_J^T\msC_J^{-1}\left(\tilde{\bPi}_J+4\msD_J^T\msZ^{-1}\tilde{\bP}\right)\right]+U(\tilde{\bPhi}_J),
	\end{align}
	where we have defined vectors $\tilde{\bY}^T=(\tilde{\bQ}^T,\tilde{\bPhi}_c^T,\tilde{\bP}^T,\tilde{\bPi}_c^T)$, and $\tilde{\bX}^T=(\tilde{\bQ}^T,\tilde{\bPhi}_c^T,\tilde{\bP}^T,\tilde{\bPi}_c^T,\tilde{\bPhi}_J^T,\tilde{\bPi}_J^T)$, the nonlinear potential  $U(\bPhi_J)=-\sum_n E_{Jn}\cos(\varphi_{Jn})$, with the matrices 
	\begin{align}
		\msH=&\begin{pmatrix}
			\msH_{\text{lin}}&\msH_{\text{int}}\\\msH_{\text{int}}^T&\msH_{J}
		\end{pmatrix},\label{eq:Hmat_JJs_Cc_Ymat_purely_quadratic}\\
		\msH_{\text{lin}}=&\begin{pmatrix}
			\frac{\msZ^T\msL^{-1}\msZ}{4}&\frac{-\msZ^T\msL^{-1}\msD_c}{2}&\frac{-\msZ^T\msL^{-1}}{2}&0\\
			\frac{-\msD_c^T\msL^{-1}\tilde{\msZ}}{2}&\msD_c^T\msL^{-1}\msD_c&\msD_c^T\msL^{-1}&0\\
			\frac{-\msL^{-1}\msZ}{2}&\msL^{-1}\msD_c&\tilde{\msL}^{-1}&0\\
			0&0&0&\msC_c^{-1}
		\end{pmatrix},\label{eq:H_lin_app}
	\end{align}
where $\tilde{\msL}^{-1}=\msL^{-1}-4\msZ^{-1}\msD_J\msC_J^{-1}\msD_J^T\msZ^{-1}$. Note that the Josephson variables $\tilde{\bPhi}_J$ only appear in the nonlinear potential, and as previously mentioned, they should be modeled with compact variables $\varphi_{Jn}=(2\pi)\frac{\Phi_{Jn}}{\Phi_0}\in\mcl{S}^1$. We then perform a rescaling transformations (another trivial transformation) so that all classical coordinates have the same units,
	\begin{align}\label{eq:rescale}
		\bar{\bQ}&=\sqrt{R}\tilde{\bQ},\qquad\bar{\bP}=\tilde{\bP}/\sqrt{R},\\
		\bar{\bPhi}_{c/J}&=\msC_{c/J}^{-\frac{1}{4}}\msL^{\frac{1}{4}}\tilde{\bPhi}_{c/J},\qquad	\bar{\bPi}_{c/J}=\msC_{c/J}^{\frac{1}{4}}\msL^{-\frac{1}{4}}\tilde{\bPi}_{c/J}.\nonumber
	\end{align}
	The matrix of Hamiltonian (\ref{eq:H_JJs_Cc_Ymat}) can now be written, for the specific  example at hand, Eq. \eqref{eq:example}, in terms of three frequencies,  $\Omega_{J}=1/\sqrt{C_J L}$, $\Omega_c=1/\sqrt{C_c L}$, and $\Omega=R/L$. Furthermore, without loss of generality, we set the transformer turn-ratios matrix to $n_{11}=n_{12}=n_{22}=1$ and $n_{21}=0$.

	\subsection{No Josephson junctions}
	Let us begin the analysis of the circuit by setting to zero the nonlinear potential, i.e., $E_{Jn}\rightarrow0$ such that $U(\bPhi_J)=0$ in Hamiltonian (\ref{eq:H_JJs_Cc_Ymat}). In the example, Eq. \eqref{eq:example},  after the  transformations of Eq. \eqref{eq:rescale} and using the  transformer turn-ratios mentioned above, the Hamiltonian in this case becomes $2H_{U=0}=\bar{\bX}^T\msH\bar{\bX}$, with 
	
	\begin{align}
		\msH=\left(
		\begin{array}{cccccccccccc}
			\frac{\Omega }{4} & 0 & \frac{\sqrt{\Omega \Omega_c}}{2} & \frac{\sqrt{\Omega \Omega_c}}{2} & 0 & -\frac{\Omega }{2} & 0 & 0 & 0 & 0 & 0 & 0 \\
			0 & \frac{\Omega }{4} & -\frac{\sqrt{\Omega \Omega_c}}{2} & 0 & \frac{\Omega }{2} & 0 & 0 & 0 & 0 & 0 & 0 & 0 \\
			\frac{\sqrt{\Omega \Omega_c}}{2} & -\frac{\sqrt{\Omega \Omega_c}}{2} & 2\Omega_c &\Omega_c & -\sqrt{\Omega \Omega_c} & -\sqrt{\Omega \Omega_c} & 0 & 0 & 0 & 0 & 0 & 0 \\
			\frac{\sqrt{\Omega \Omega_c}}{2} & 0 &\Omega_c &\Omega_c & 0 & -\sqrt{\Omega \Omega_c} & 0 & 0 & 0 & 0 & 0 & 0 \\
			0 & \frac{\Omega }{2} & -\sqrt{\Omega \Omega_c} & 0 & \frac{8\Omega_J^2}{\Omega }+\Omega  & -\frac{4\Omega_J^2}{\Omega } & 0 & 0 & 0 & 0 & \frac{2\Omega_J^{3/2}}{\sqrt{\Omega }} & \frac{2\Omega_J^{3/2}}{\sqrt{\Omega }} \\
			-\frac{\Omega }{2} & 0 & -\sqrt{\Omega \Omega_c} & -\sqrt{\Omega \Omega_c} & -\frac{4\Omega_J^2}{\Omega } & \frac{4\Omega_J^2}{\Omega }+\Omega  & 0 & 0 & 0 & 0 & -\frac{2\Omega_J^{3/2}}{\sqrt{\Omega }} & 0 \\
			0 & 0 & 0 & 0 & 0 & 0 &\Omega_c & 0 & 0 & 0 & 0 & 0 \\
			0 & 0 & 0 & 0 & 0 & 0 & 0 &\Omega_c & 0 & 0 & 0 & 0 \\
			0 & 0 & 0 & 0 & 0 & 0 & 0 & 0 & 0 & 0 & 0 & 0 \\
			0 & 0 & 0 & 0 & 0 & 0 & 0 & 0 & 0 & 0 & 0 & 0 \\
			0 & 0 & 0 & 0 & \frac{2\Omega_J^{3/2}}{\sqrt{\Omega }} & -\frac{2\Omega_J^{3/2}}{\sqrt{\Omega }} & 0 & 0 & 0 & 0 &\Omega_J & 0 \\
			0 & 0 & 0 & 0 & \frac{2\Omega_J^{3/2}}{\sqrt{\Omega }} & 0 & 0 & 0 & 0 & 0 & 0 &\Omega_J \\
		\end{array}
		\right)
	\end{align}
	and $\bar{\bX}^T=(\bar{\bQ}^T,\bar{\bPhi}_c^T,\bar{\bP}^T,\bar{\bPi}_c^T,\bar{\bPhi}_J^T,\bar{\bPi}_J^T)$. Solving the eigenvalue problem for $\msJH$ we can find the symplectic transformation.  For the homogeneous case of frequencies $\Omega=\Omega_c=\Omega_J=1$, say, we have
	\begin{align}
		\msS|_{\Omega_\alpha=1}\stackrel{\text{num.}}{=}&\left(
		\begin{array}{cccccccccccc}
			-1 & 1 & -1 & -1 & \sqrt{2} & \frac{1}{\sqrt{2}} & 0 & 0 & 1.38612 & 0.435045 & 0. & 0. \\
			1 & 0 & 0 & 1 & -\sqrt{2} & 0 & 0 & 0 & -0.634047 & 0.634047 & 0.800274 & -0.251173 \\
			0 & 0 & 0 & \frac{1}{2} & -\frac{1}{\sqrt{2}} & 0 & 0 & 0 & 0.317023 & -0.317023 & -0.125587 & 0.400137 \\
			0 & 0 & \frac{1}{2} & 0 & 0 & -\frac{1}{2 \sqrt{2}} & 0 & 0 & 0.217523 & 0.693058 & 0. & 0. \\
			-\frac{1}{2} & 0 & 0 & 0 & 0 & 0 & 0 & 0 & 0. & 0. & 0.27455 & 0.27455 \\
			-\frac{1}{2} & \frac{1}{2} & 0 & 0 & 0 & 0 & 0 & 0 & -0.158512 & 0.158512 & -0.125587 & 0.400137 \\
			0 & 0 & 0 & 0 & 0 & 0 & -\frac{1}{\sqrt{2}} & 0 & 0.317023 & -0.317023 & 0.800274 & -0.251173 \\
			0 & 0 & 0 & 0 & 0 & 0 & 0 & -\frac{1}{\sqrt{2}} & 0. & 0. & 0.5491 & 0.5491 \\
			0 & 0 & 0 & \frac{1}{2} & \frac{1}{\sqrt{2}} & 0 & 0 & 0 & 0.317023 & -0.317023 & -0.125587 & 0.400137 \\
			0 & 0 & \frac{1}{2} & 0 & 0 & \frac{1}{2 \sqrt{2}} & 0 & 0 & 0.217523 & 0.693058 & 0. & 0. \\
			0 & 1 & 0 & 0 & 0 & 0 & \frac{1}{\sqrt{2}} & 0 & 0. & 0. & 0. & 0. \\
			1 & 0 & 0 & 0 & 0 & 0 & 0 & \frac{1}{\sqrt{2}} & 0. & 0. & 0. & 0. \\
		\end{array}
		\right)
	\end{align}
\end{widetext}
	such that the diagonalized Hamiltonian has the form (\ref{eq:sthsexplicit}) where the matrix of frequencies is 
	\begin{align}
		\msOmega=\begin{pmatrix}
			2.52434 & 0\\
			0& 0.792287 \\
		\end{pmatrix}.
	\end{align}
	As previously advanced in the analysis of the main text, this system has two free-particles (which correspond to the fact that we have two pairs of capacitors in series and we have used a redudant description), two nondynamical pairs of conjugated variables (associated with the intrinsically redundant gyrator description), and two harmonic oscillators. 
	\subsection{Keeping the full cosine}
	Let us now analyse the full problem with the Josephson junctions. One of the routines suggested for the quantization of this problem is to perform Williamson's analysis only on the fully linear subsector, in this case, $\msH_{\text{lin}}$ of Eq.  (\ref{eq:H_lin_app}), i.e., taking into consideration only the inductor and coupling capacitor variables. This partial classical diagonalization will be followed by a quantum one on the coupled quantized system, thus the name two tier method. The partial symplectic diagonalization of the Hamiltonian yields
	\begin{align}
		\msH_{pD}=\msS_{\bY}^T\msH\msS_{\bY}&=\begin{pmatrix}
			0&0&0&0&0\\
			0&\msOmega&0&0&\msG\\
			0&0&\msOmega&0&\msM\\
			0&0&0&0&0\\
			0&\msG^T&\msM^T&0&\msOmega_J
		\end{pmatrix},
	\end{align}
with particular numbers for the set of parameters chosen
	\begin{widetext}
\begin{align}
	\msH_{pD}\stackrel{num}{=}\left(
		\begin{array}{cccccccccccc}
			0& 0& 0& 0& 0& 0& 0& 0& 0& 0& 0& 0\\
			0& 0& 0& 0& 0& 0& 0& 0& 0& 0& 0& 0\\
			0& 0& 0& 0& 0& 0& 0& 0& 0& 0& 0& 0\\
			0& 0& 0& 0& 0& 0& 0& 0& 0& 0& 0& 0\\
			0& 0& 0& 0& 2.52434 & 0& 0& 0& 0& 0& 0.800274 & 0.5491 \\
			0& 0& 0& 0& 0& 0.792287 & 0& 0& 0& 0& -0.251173 & 0.5491 \\
			0& 0& 0& 0& 0& 0& 2.52434 & 0& 0& 0& -0.317023 & 0\\
			0& 0& 0& 0& 0& 0& 0& 0.792287 & 0& 0& 0.317023 & 0\\
			0& 0& 0& 0& 0& 0& 0& 0& 0& 0& 0& 0\\
			0& 0& 0& 0& 0& 0& 0& 0& 0& 0& 0& 0\\
			0& 0& 0& 0& 0.800274 & -0.251173 & -0.317023 & 0.317023 & 0& 0& 1. & 0\\
			0& 0& 0& 0& 0.5491 & 0.5491 & 0& 0& 0& 0& 0& 1. \\
		\end{array}
		\right),
	\end{align}
\end{widetext}
where the coordinates have been transformed accordingly to the partially-diagonalizing symplectic transformation such that  $(\msS_{\bY}^{-1}\bar{\bX})^T=(\bar{\by}^T,\bar{\bpi}^T,\bar{\bPhi}_J^T,\bar{\bPi}_J^T)$, and $\bar{\by}^T=(\bar{\by}_{\mathrm{nd}}^T,\bar{\by}_{\mathrm{ho}}^T)$ are diagonalized {\it position} coordinates conjugated to the canonical {\it momenta} $\bar{\bpi}^T=(\bar{\bpi}_{\mathrm{nd}}^T,\bar{\bpi}_{\mathrm{ho}}^T)$. After the canonical quantization of purely oscillator conjugated pairs and their transformation to annihilation and creation operators, i.e.,  $\bar{y}_i=\sqrt{\frac{\hbar}{2}}(a_i+a_i^\dag)$, and $\bar{\pi}_i=i\sqrt{\frac{\hbar}{2}}(a_i^\dag-a_i)$, we obtain
\begin{align}
	\hat{H}/\hbar\stackrel{\text{q.}}{=}&\sum_{n,m}^2\left[\Omega_n a_n^\dag a_n + (g_{nm}a_n+g_{nm}^*a_n^\dag)\hat{\bar{\Pi}}_{Jm}\right]+\hat{H}_{J}/\hbar,\label{eq:H_2port_Ymat_Cc_JJs_separated_app}
\end{align}
where $g_{nm}=\sqrt{\frac{1}{2\hbar}}\left[\msG_{nm}-i\msM_{nm}\right]$. The purely quadratic part of the Josephson Hamiltonian can be read from the lower right corner $4\times4$ submatrix of $\msH_{pD}$ (or $\msH$), and thus the Josephson part of the total Hamiltonian becomes
\begin{align}
	\hat{H}_J/\hbar=&\sum_{n=1}^2\hat{\bar{\Pi}}_{Jn}^2-E_{Jn}\cos(\hat{\bar{\varphi}}_{Jn}),\label{eq:H_J_separated_app}
\end{align}
where the renormalized phase variable is  $\bar{\varphi}_{Jl}=2\pi\frac{\bar{\Phi}_{Jl}}{\Phi_0}$.
\subsection{Black-box quantization approach}
Let us now present a  black-box quantization approach. Under the condition that the Josephson flux variables have small fluctuations, we carry out an extended harmonic analysis. To this point, extract the quadratic term of the  cosine potentials, and include its contribution, $\bar{\bPhi}_J^T\msL_J^{-1}\bar{\bPhi}_J$, to the Hamiltonian matrix (\ref{eq:Hmat_JJs_Cc_Ymat_purely_quadratic}), with
\begin{align}
	\qquad\qquad\msL_J^{-1}=&\sqrt{\frac{L}{L_J}}\Omega_J\mone_2.\label{eq:L_J_bb}
\end{align}

Setting $L_J$, as way of illustration, to be equal to  $L$, one would obtain now four independent harmonic dynamics (the two free particles now are trapped in confining potentials) whereas the nondynamical sector remains unchanged (two pairs of conjugated variables) such that the final canonical block diagonal matrix $\msH_D=\msS^T\msH\msS$ (Eq.  (\ref{eq:sthsexplicit})) has for its frequency block matrix
\begin{align}
	\msOmega=\begin{pmatrix}
		2.61227 & 0& 0& 0\\
		0& 1.20569 & 0& 0\\
		0& 0& 0.73029 & 0\\
		0& 0& 0& 0.434761 \\
	\end{pmatrix}.
\end{align}
The final black-box Hamiltonian, after quantization and transformation to annihilation ($b_n$) and creation ($b_n^\dag$) operators, reads
\begin{align}
	\hat{H}_{bb}\stackrel{\text{q.}}{=}\sum_{n}^4\hbar\Omega_n b_n^\dag b_n+E_{J}\sum_{l={1,2}}\sum_{m=2}^\infty\frac{(-1)^{m+1}\hat{\bar{\varphi}}_{Jl}^{2m}}{(2m)!}.
\end{align}
The frequencies $\Omega_n$ can also be obtained by going through  Hamiltonian (\ref{eq:H_2port_Ymat_Cc_JJs_separated_app}), by separating out the quadratic part of $H_J$, Eq. (\ref{eq:H_J_separated_app}), and carrying out symplectic diagonalization of the small oscillations approximation. Naturally enough, the results match.

As already mentioned in the main text, in this specific case the  rescaled Josephson coordinates do not depend on the nondynamical sector of the Williamson diagonalization. This statement follows from the explicit form of  the diagonalizing symplectic transformation in the black-box approach, i.e., $\msS\bZ=\bar{\bX}$
\begin{align}
	\msS=\left(
	\begin{array}{cccccccccccc}
		-1 & -1 & 1 & -2 & *&*&*&*&*&*&*&*\\
		1 & 1 & 0 & 0 &  *&*&*&*&*&*&*&*\\
		0 & 1 & 0 & 0 &  *&*&*&*&*&*&*&*\\
		0 & 0 & 0 & 1 &  *&*&*&*&*&*&*&*\\
		-\frac{1}{2} & \frac{1}{2} & 0 & 0 &  *&*&*&*&*&*&*&*\\
		-\frac{1}{2} & \frac{1}{2} & \frac{1}{2} & 0 &  *&*&*&*&*&*&*&*\\
		0 & 0 & 0 & 0 &  *&*&*&*&*&*&*&*\\
		0 & 0 & 0 & 0 &  *&*&*&*&*&*&*&*\\
		0 & 0 & 0 & 0 &   *&*&*&*&*&*&*&*\\
		0 & 0 & 0 & 0 &  *&*&*&*&*&*&*&*\\
		0 & 0 & 1 & 0 &  *&*&*&*&*&*&*&*\\
		1 & -1 & 0 & 0 &  *&*&*&*&*&*&*&*\\
	\end{array}
	\right),
\end{align}
where the ninth and tenth row have zeroes in the first four columns, meaning that $\bar{\varphi}_{Ji}(\bz_{ho},\bpi_{ho})$, as $\bZ^T=(\bsb{z}_{nd}^T,\bpi_{nd}^T,\bz_{ho}^T,\bpi_{ho}^T)$. 

However, it must be remarked that this could  be not the case in generic networks with Josephson junctions and ideal nonreciprocal elements. In a more general scenario, e.g., by removing the coupling capacitors $C_{ci}$ in circuit of Fig.~\ref{fig:2CQ_2_port_NR_Y}, the Josephson variables would mix with conjugate position and momenta  variables that are nondynamical in the linear approximation,  but change character for the full problem, i.e., we would have a dependence of the  (rescaled) Josephson variables of the form  $ \bar{\varphi}_{Ji}(\bz_{nd}^T,\bpi_{nd}^T,\bz_{ho}^T,\bpi_{ho}^T)$, in such a way  that the Hamiltonian would be presented as
\begin{align}
	H=H_{ho}(\bz_{ho}^T,\bpi_{ho}^T)+ H_{nl}(\bz_{nd}^T,\bpi_{nd}^T,\bz_{ho}^T,\bpi_{ho}^T). 
\end{align}
As both Williamson nondynamical positions and momenta  are present, their classical equations of motion are no longer trivial, and they become dynamical.
Thus, the black-box approach would  provide a starting point  for the  study of  the low-lying energy sector of such classes of circuits, but would require  refinement additional to the symplectic diagonalization step we put forward.

\addcontentsline{toc}{section}{References}
\bibliography{bibliography}

\end{document}